\documentclass{scrartcl} 
 
\usepackage{amsmath,latexsym,amsopn,amsfonts} 
\usepackage{algorithmic}
\usepackage{graphicx,fullpage}
\usepackage{textcomp}
\usepackage{xcolor,url}
\def\BibTeX{{\rm B\kern-.05em{\sc i\kern-.025em b}\kern-.08em
		T\kern-.1667em\lower.7ex\hbox{E}\kern-.125emX}}

\usepackage[boxed,ruled,vlined,linesnumbered,noresetcount]{algorithm2e}
\usepackage{mdwlist}
\usepackage{graphicx,xspace}

\newtheorem{theorem}{Theorem}[section]
\newenvironment{Theorem}{\begin{theorem}}{\end{theorem}} 
\newtheorem{lemma}[theorem]{Lemma}
\newenvironment{Lemma}{\begin{lemma}}{\end{lemma}} 
\newtheorem{definition}{Definition}[section]
\newenvironment{Definition}{\begin{definition}}{\end{definition}} 

\newcommand\bull{{\operatorname{-\xspace}}}
\newcommand{\SectionAbv}{Section\xspace}
\newcommand{\SectionsAbv}{Sections\xspace}

\usepackage{ stmaryrd }
\newcommand{\blitza}{\lightning}

\usepackage{color}
\newcommand{\ems}[1]{\textcolor{black}{#1}}
\newcommand{\emsA}[1]{\textcolor{black}{#1}}
\newcommand{\emsB}[1]{\textcolor{black}{#1}}
\newcommand{\emsC}[1]{\textcolor{black}{#1}}
\newcommand{\emsD}[1]{\textcolor{black}{#1}}
\newcommand{\emsE}[1]{\textcolor{black}{#1}}
\newcommand{\emsF}[1]{\textcolor{black}{#1}}
\newcommand{\rd}[1]{\textcolor{black}{#1}}
\newcommand{\rdA}[1]{\textcolor{black}{#1}}
\newcommand{\bigO}{\mathcal{O}\xspace}
\newcommand{\remove}[1]{}
\newcommand{\reduce}[1]{#1}
\newcommand{\technicalReport}[1]{#1}
\newcommand{\extendedAbstract}[1]{}

\newcommand{\Correct}{\mathit{Correct}\xspace} 
\newcommand{\init}{\texttt{init}\xspace} 
\newcommand{\initE}{\emph{\texttt{init}}\xspace} 
\newcommand{\validE}{\emph{\texttt{valid}}\xspace} 
\newcommand{\echo}{\texttt{echo}\xspace} 
\newcommand{\ready}{\texttt{ready}\xspace} 
\newcommand{\valid}{\texttt{valid}\xspace} 
\newcommand{\etal}{\emph{et al.}\xspace}
\newcommand{\eg}{\emph{e.g.,}\xspace}
\newcommand{\Eg}{\emph{E.g.,}\xspace}
\newcommand{\ie}{\emph{i.e.,}\xspace}
\newcommand{\Ie}{\emph{I.e.,}\xspace}

\newcommand{\true}{\mathsf{True}\xspace}
\newcommand{\True}{\textsf{True}\xspace}
\newcommand{\false}{\mathsf{False}\xspace}
\newcommand{\False}{\textsf{False}\xspace}
\newcommand{\sP}{\mathcal{P}\xspace}
\newcommand{\bZ}{{Z}\xspace} 

\newcommand{\capacity}{\mathsf{channelCapacity}\xspace} 
\newcommand{\done}{\mathsf{result}\xspace}
\newcommand{\bcdone}{\mathsf{result}\xspace}
\newcommand{\sameValue}{\mathit{sameValue}\xspace}
\newcommand{\binValues}{\mathit{binValues}\xspace}
\newcommand{\Figure}{Fig.\xspace}
\newcommand{\typ}{\emsA{\mathit{phs}\xspace}}
\newcommand{\wrt}{\emph{w.r.t.}\xspace}
\newcommand{\hasTerminated}{\mathsf{hasTerminated}\xspace}

\newcommand{\respectively}{resp.\xspace}
\newcommand{\respectivelyC}{resp.,\xspace}
\newcommand{\respectivelyP}{resp.\xspace}

\newenvironment{lemmaProof}{\par\noindent\textbf{Proof of Lemma  \lemcnt\space}}{\hfill $\Box_{Lemma ~ \lemcnt}$}

\newenvironment{theoremProof}{\par\noindent\textbf{Proof of Theorem  \thmcnt\space}}{\hfill $\Box_{Theorem ~ \thmcnt}$}

\newcommand{\clmcnt}{0}
\newcommand{\lemcnt}{0}
\newcommand{\thmcnt}{0}

\newcommand{\Section}[1]{\section{#1}}
\newcommand{\Subsection}[1]{\subsection{#1}}
\newcommand{\Subsubsection}[1]{\subsubsection{#1}}

\newcommand{\algSize}{normalsize} 

\date{}
\begin{document}
	
	\title{\LARGE{Self-stabilizing Byzantine Multivalued Consensus}~\footnote{The extended abstract version of this work appear in~\cite{proceedings}.}\\\Large{(technical report)}}

	
	\author{Romaric Duvignau~\footnote{Chalmers University of Technology, Sweden. 	Email: \texttt{\{duvignau,elad\}@chalmers.se}}~ Michel Raynal~\footnote{IRISA, University Rennes 1, France. Email: \texttt{michel.raynal@irisa.fr}} and Elad M.\ Schiller~\footnotemark[2]}

	\maketitle
	
	\begin{abstract}
		Consensus, abstracting a myriad of problems in which processes have to agree on a single value, is one of the most celebrated problems of fault-tolerant distributed computing. Consensus applications include fundamental services for the environments of the Cloud and Blockchain, and in such challenging environments, malicious behaviors are often modeled as adversarial Byzantine faults. 
		
		At OPODIS 2010, Mostéfaoui and Raynal (in short MR) presented a Byzantine-tolerant solution to consensus in which the decided value cannot be a value proposed only by Byzantine processes. MR has optimal resilience coping with up to $t < n/3$ Byzantine nodes over $n$ processes. MR provides this multivalued consensus object (which accepts proposals taken from a finite set of values) assuming the availability of a single Binary consensus object (which accepts proposals taken from the set $\{0,1\}$).
		
		\emsD{This work, which focuses on multivalued consensus, aims at the design of an even more robust solution than MR. Our proposal expands MR's fault-model with self-stabilization, a vigorous notion of fault-tolerance. In addition to tolerating Byzantine, self-stabilizing systems can automatically recover after the occurrence of \emph{arbitrary transient-faults}. These faults represent any violation of the assumptions according to which the system was designed to operate (provided that the algorithm code remains intact).}
		
		To the best of our knowledge, we propose the first self-stabilizing solution for intrusion-tolerant multivalued consensus for asynchronous message-passing systems prone to Byzantine failures. 
		\emsE{Our} solution has a $\bigO(t)$ stabilization time from arbitrary transient faults.
	\end{abstract}

	

	\Section{Introduction}
	\label{sec:intro}
	We present in this work a novel self-stabilizing algorithm for multivalued consensus in signature-free asynchronous message-passing systems that can tolerate Byzantine faults. 
	We provide rigorous correctness proofs to demonstrate that our solution is correct and outperforms all previous approaches in terms of its fault tolerance capabilities, and further analyze its \emsE{recovery} time.
	Compared to existing solutions, our proposed algorithm represents a significant advancement in the state of the art, as it can effectively handle a wider range of faults, including both benign and malicious failures, as well as arbitrary, transient, and possibly unforeseen violations of the assumptions according to which the system was designed to operate. 
	Our proposed solution can hence further facilitate the design of new fault-tolerant building blocks for distributed systems.  
	
	\Subsection{Task Requirements and Fault Models} 
	\label{sec:backgroundMotivation}
	\paragraph*{Multivalued Consensus (MVC)}
	The consensus problem is one of the most challenging tasks in fault-tolerant distributed computing. The problem definition is rather simple. It assumes that each non-faulty process advocates for a single value from a given set $V$. The problem of \emph{Byzantine-tolerant Consensus} (BC) requires \emph{BC-completion} (\textbf{R1}), \ie all non-faulty processes decide a value, \emph{BC-agreement} (\textbf{R2}), \ie no two non-faulty processes can decide different values, and \emph{BC-validity} (\textbf{R3}), \ie if all non-faulty processes propose the same value $v\in V$, only $v$ can be decided. When the set, $V$, from which the proposed values are taken is $\{0,1\}$, the problem is called Binary consensus and otherwise, MVC. 
	We study MVC solutions that assume access to a single Binary consensus object. 
	
	
	\paragraph*{Byzantine fault-tolerance (BFT)}
	%
	\label{sec:BYZnoIntro}
	Lamport \etal~\cite{DBLP:journals/toplas/LamportSP82} say that a process commits a \textit{Byzantine} failure if it deviates from the algorithm instructions, say, by deferring or omitting messages that were sent by the algorithm or sending fake messages, which the algorithm never sent. 
	Such malicious behaviors include crashes and may be the result of hardware or software malfunctions as well as coordinated malware attacks. 
	In order to safeguard against such attacks, Mostéfaoui and Raynal~\cite{DBLP:conf/opodis/MostefaouiR10}, MR from now on, suggested the \emph{BC-no-intrusion} (\textbf{R4}) validity requirement (aka \emph{intrusion-tolerance}). 
	Specifically, the decided value cannot be a value that was proposed \emph{only} by faulty processes. 
	Also, when it is not possible to decide on a value, the error symbol $\blitza$ is returned instead. 
	For the sake of deterministic solvability~\cite{DBLP:journals/jacm/DworkLS88,DBLP:journals/toplas/LamportSP82,DBLP:journals/jacm/PeaseSL80,DBLP:conf/srds/Perry84}, we assume that there are at most $t<n/3$ Byzantine processes in the system, where $n$ is the total number of processes. It is also well-known that no deterministic (multivalued or Binary) consensus solution exists for asynchronous systems in which at least one process may crash (or be Byzantine)~\cite{DBLP:journals/jacm/FischerLP85}. 
	Our \emsD{self-stabilizing MVC algorithm circumvents} this impossibility by assuming that the system is enriched with a \emsD{Binary consensus object,} as in the studied \emsD{(non-self-stabilizing)} solution by MR~\cite{DBLP:conf/opodis/MostefaouiR10}, \ie reducing MVC to Binary consensus.
	
	\begin{Definition}
		\label{def:MVC}
		The \textbf{BFT Multivalued Consensus} (MVC) problem requires BC-completion (R1), BC-agreement (R2), BC-validity (R3), and BC-no-Intrusion (R4). 
	\end{Definition}
	
	\paragraph*{Self-stabilization}
	%
	\label{sec:selfDisc}
	We study an asynchronous message-passing system that has no guarantees on the communication delay and the algorithm cannot explicitly access the local clock. Our fault model includes undetectable Byzantine failures.
	In addition, we aim to recover from \emph{arbitrary transient-faults}, \ie any temporary violation of assumptions according to which the system was designed to operate. 
	This includes the corruption of control variables, such as the program counter and message payloads, as well as operational assumptions, such as that there are more than $t$ faulty processes. 
	We note that non-self-stabilizing BFT systems do not consider recovery after the occurrence of such violations.
	Since the occurrence of these failures can be arbitrarily combined, we assume these transient-faults can alter the system state in unpredictable ways. 
	In particular, when modeling the system, Dijkstra~\cite{DBLP:journals/cacm/Dijkstra74} assumes that these violations bring the system to an arbitrary state from which a \emph{self-stabilizing system} should recover~\cite{DBLP:series/synthesis/2019Altisen,DBLP:books/mit/Dolev2000}. 
	\Ie Dijkstra requires (i) recovery after the last occurrence of a transient-fault and (ii) once the system has recovered, it must never violate the task requirements. 
	Arora and Gouda~\cite{DBLP:journals/tse/AroraG93} refer to the former requirement as \emph{Convergence} and the latter as \emph{Closure}.
	
	\begin{Definition}
		\label{def:SSBFT}
		A \textbf{Self-Stabilizing Byzantine Fault-Tolerant} (SSBFT) \emsD{MVC algorithm satisfies} the requirements of Definition~\ref{def:MVC} \emsD{within} the execution of a finite number of steps following the last transient fault, \emsD{which} left the system in an arbitrary state.
	\end{Definition}

	\Subsection{Related work}
	Ever since the\reduce{ seminal} work of Lamport, Shostak, and Pease~\cite{DBLP:journals/toplas/LamportSP82}\reduce{ four decades ago}, BFT consensus has been an active research subject, see~\cite{DBLP:journals/ijccbs/CorreiaVNV11} and references therein. The recent rise of distributed ledger technologies, \eg~\cite{DBLP:conf/sp/AbrahamMN0Y20}, brought phenomenal attention to the subject. 
	\emsD{We aim to provide a degree of dependability that is higher than existing solutions.}
	
	Ben-Or, Kelmer, and Rabin~\cite{DBLP:conf/podc/Ben-OrKR94} were the first to show that BFT MVC can be reduced to Binary consensus. 
	Correia, Neves, and Ver{\'{\i}}ssimo~\cite{DBLP:journals/cj/CorreiaNV06,DBLP:journals/tpds/NevesCV05} later established the connection between intrusion tolerance and Byzantine resistance.
	These ideas form the basis of the MR algorithm~\cite{DBLP:conf/opodis/MostefaouiR10}.
	MR is a leaderless consensus algorithm~\cite{DBLP:conf/icdcs/AntoniadisDGGZ21} and as such, it avoids the key weakness of leader-based algorithms~\cite{DBLP:journals/tocs/CastroL02} when the leader is slow and delays termination.
	There exist self-stabilizing solutions for MVC but only crash-tolerant ones~\cite{DBLP:conf/netys/BlanchardDBD14,DBLP:journals/jcss/DolevKS10,DBLP:conf/edcc/LundstromRS21,DBLP:conf/icdcn/LundstromRS21,DBLP:conf/netys/GeorgiouLS19}, whereas, the existing BFT solutions are not self-stabilizing~\cite{DBLP:books/sp/Raynal18}.   
	For example, the recent self-stabilizing crash-tolerant \emsD{MVC in~\cite{DBLP:conf/edcc/LundstromRS21} solves a less challenging problem than the SSBFT problem studied} here since it does not account for malicious behaviors.
	Mostéfaoui, Moumen, and Raynal~\cite{DBLP:conf/podc/MostefaouiMR14} {(MMR in short)} presented BFT algorithms for solving Binary consensus using common coins, of which 
	~\cite{DBLP:conf/netys/GeorgiouMRS21,DBLP:conf/sss/GeorgiouRS23} recently introduced a self-stabilizing variation that satisfies the safety requirements, \ie agreement and validity, with an exponentially high probability that depends only on a predefined constant, \emsD{which safeguards safety.}
	%
	%
	\emsF{The} related work \emsD{also} includes SSBFT state-machine replication by Binun \etal~\cite{DBLP:conf/sss/BinunCDKLPYY16,DBLP:conf/cscml/BinunDH19} for synchronous systems and Dolev \etal~\cite{DBLP:conf/cscml/DolevGMS18} for practically-self-stabilizing partially-synchronous systems. \emsF{Note that both Binun \etal and Dolev \etal study another problem for another kind of system setting.
		In~\cite{DBLP:conf/sss/DolevLS13}, the problems of SSBFT topology discovery and message
		delivery were studied. Self-stabilizing atomic memory under semi-Byzantine adversary is studied in~\cite{DBLP:journals/algorithmica/DolevPS23}.
	} 
	
	\begin{figure}
		\begin{center}
			\includegraphics[scale=0.4, clip]{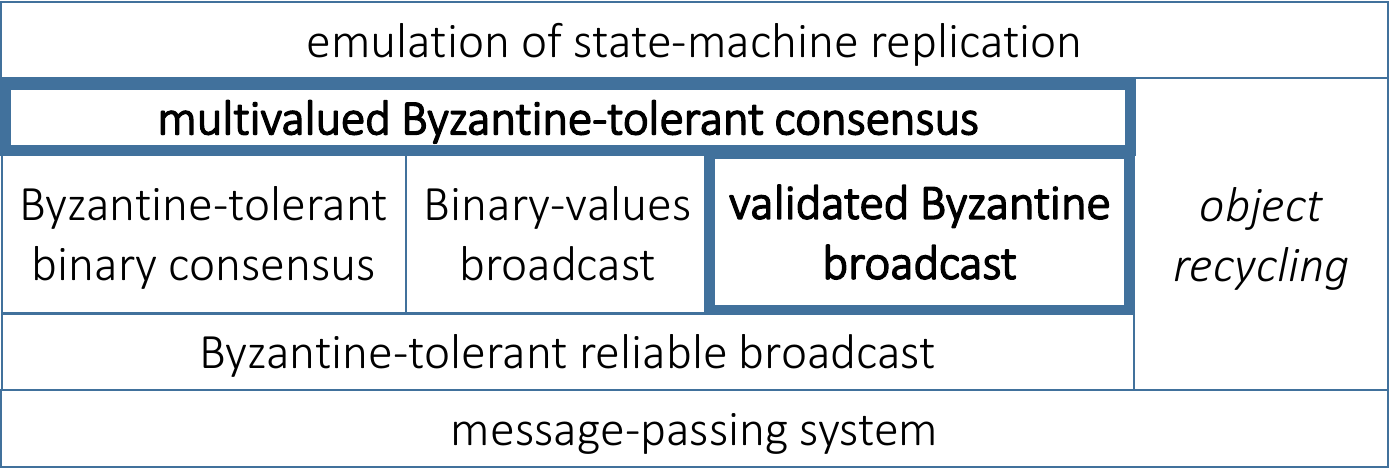}
		\end{center}
		\caption{\label{fig:suit}\emsE{We assume the availability of SSBFT protocols \rdA{(cf. Definition~\ref{def:SSBFT})} for Binary consensus and object recycling. The studied problems appear in boldface fonts. The other layers, BRB, BV-broadcast, and state machine replication, are in plain font.}}
		%
	\end{figure}
	
	\Subsection{A brief overview of the MR algorithm}
	\label{sec:OverMRAlgo}
	\emsD{The MR algorithm assumes that all (non-faulty) processes eventually propose a value.
		Upon the proposal of value $v$, the algorithm utilizes a Validated Byzantine Broadcast protocol, known as VBB, to enable each process to reliably deliver $v$. 
		The VBB-delivered value could be either the message, $v$, which was VBB-broadcast, or $\bot$ when $v$ could not be validated.
		For a value $v$ to be valid, it must be VBB-broadcast by at least one non-faulty process.}
	
	Following the VBB-delivery from at least $n-t$ different processes, MR undergoes a local test,\reduce{ which is} detailed in \SectionAbv~\ref{sec:MVC}.
	If at least one non-faulty process passes this test, it implies that all non-faulty processes can ultimately agree on a single value proposed by at least one non-faulty process.
	Therefore, the MR algorithm employs Byzantine-tolerant Binary consensus to reach consensus on the outcome of the local test. 
	If the agreed value indicates the existence of at least one non-faulty process that has passed the test, then each non-faulty process waits until it receives at least $n-2t$ VBB-arrivals with the same value, $v$, which is the decided value in this instance of multivalued consensus. 
	If no such indication is represented by the agreed value, the MR algorithm reports its inability to decide in this MVC invocation.     
	For further information, please refer to Section~\ref{sec:algos}.
	
	\Subsection{\emsD{Our SSBFT variation on MR}}
	\label{sec:ourSSBFTvarMR}
	\emsD{This work considers transformers that take algorithms as input and output their self-stabilizing variations. 
		For example, Duvignau, Raynal, and Schiller~\cite{DBLP:journals/corr/abs-2201-12880} (referred to as DRS) proposed a transformation for converting the Byzantine Reliable Broadcast (referred to as BRB) algorithm, originally introduced by Bracha and Toueg~\cite{DBLP:conf/podc/BrachaT83}, into a Self-Stabilizing BFT (in short, SSBFT) variation. 
		Another transformation, proposed by Georgiou, Marcoullis, Raynal, and Schiller~\cite{DBLP:conf/netys/GeorgiouMRS21,DBLP:conf/sss/GeorgiouRS23} (referred to as GMRS), presented the SSBFT variation of the BFT Binary consensus algorithm of MMR.}
	
	Our transformation builds upon the works of DRS and GMRS when transforming the (non-stabilizing) BFT MR algorithm into its self-stabilizing variation.
	The design of SSBFT solutions requires addressing considerations that BFT solutions do not need to handle, as they do not consider transient faults. 
	%
	
	\emsE{For instance,}
	MR uses a (non-self-stabilizing) BFT Binary consensus object, denoted as $obj$. 
	In MR, $obj$ returns a value that is proposed by at least one non-faulty process, which corresponds to a test result (as mentioned in \SectionAbv~\ref{sec:OverMRAlgo}\reduce{ and detailed in \SectionAbv~\ref{sec:vbbInv}}). 
	However, a\reduce{ single} transient fault can change $obj$'s value from $\false$, \ie not passing the test, to $\true$.
	Such an event would cause MR, which was not designed to tolerate transient faults, to wait indefinitely for messages that are never sent.
	Our solution addresses this issue by carefully integrating GMRS's\reduce{ SSBFT} Binary-values broadcast (in short, BV-broadcast). This subroutine ensures that $obj$'s value is proposed by at least one non-faulty node even in the presence of transient faults.
	
	The vulnerability of consensus objects to corruption by transient faults holds true regardless of whether we consider Binary or multivalued consensus (MVC).
	Thus, our SSBFT MVC solution is required to decide even when starting from an arbitrary state. 
	To achieve this, our correctness proof demonstrates that our solution always terminates.
	We borrow from GMRS the concept of \emph{consensus object recycling}, which refers to reusing the object (space in the local memory of all non-faulty processes) for a later MVC invocation.
	Even when starting from an arbitrary state, the proposed solution decides on a value that is eventually delivered to all non-faulty processes, albeit potentially violating safety due to the occurrence of transient faults. 
	Then, utilizing GMRS's subroutine for recycling consensus objects, the MVC object is recycled.
	Starting from a post-recycling state, the MVC object guarantees both safety and liveness for an unbounded number of invocations.
	This is one of the principal arguments behind our correctness proof.
	
	\emsD{We clarify that GMRS's recycling subroutine relies on synchrony assumptions.
		To mitigate the impact of these assumptions, a single recycling action can be performed for a batch of $\delta$ objects, where $\delta$ is a predefined constant determined by the memory available for consensus objects.
		This approach allows for asynchronous networking in communication-intensive components, such as the consensus objects, while the synchronous recycling actions occur according to the predefined load parameter, $\delta$.}
	

	\emsF{We want to emphasize to the reader that, although our solution is built upon the previous works of DRS~\cite{DBLP:journals/corr/abs-2201-12880} and GMRS~\cite{DBLP:conf/netys/GeorgiouMRS21,DBLP:conf/sss/GeorgiouRS23} (which addressed different problems than the one under study), we encounter similar challenges in the transformation of code from non-self-stabilizing to self-stabilizing algorithms. Nevertheless, achieving the desired self-stabilizing properties in our construction necessitates a thoughtful combination of SSBFT building blocks and a meticulous analysis of the transformed algorithms.
		This combination process cannot be derived from the DRS and GMRS transformations.
		The self-stabilizing issues to tackle that are inherent to the studied algorithms are further explained in \SectionAbv~\ref{sec:vbbInv} for the VBB algorithm and in \SectionAbv~\ref{sec:mvcInv} for the MVC algorithm.}


	\Subsection{Our contribution}
	\label{sec:ourContribution}
	We present a fundamental module for dependable distributed systems: an SSBFT MVC algorithm for asynchronous message-passing systems. 
	\emsD{Hence, we advance the state of the art \wrt the dependability degree.}
	We obtain this new self-stabilizing algorithm via a transformation of the non-self-stabilizing MR algorithm.
	MR offers optimal resilience by assuming $t < n/3$, where $t$ is the number of faulty processes and $n$ is the total number of processes.
	\emsD{Our solution preserves} this optimality. 
	
	In the absence of transient faults, our solution achieves consensus within a constant \emsE{number} of communication rounds{ during arbitrary executions and} without fairness assumptions. 
	After the occurrence of any finite number of arbitrary transient faults, the system recovers within a constant number of invocations of the underlying communication abstractions.
	This implies recovery within a constant time (in terms of asynchronous cycles), assuming execution fairness among the non-faulty processes.
	We clarify that these fairness assumptions are only needed for a bounded time, \ie during recovery, and not during the period in which the system is required to satisfy the task requirements (Definition~\ref{def:MVC}).
	\emsE{It is important to note that when taking into account also the stabilization time of the underlying communication abstractions, the recycling mechanism stabilizes within $\bigO(t)$ synchronous rounds.}
	
	The communication costs of the studied algorithm, \ie MR, and \emsD{the} proposed one are similar in the number of \emsE{BRB} and Binary consensus invocations. 
	The main difference is that our SSBFT solution uses BV-broadcast for making sure that the value decided by the SSBFT Binary consensus object remains consistent until the proposed SSBFT solution completes and is ready to be recycled.
	
	To the best of our knowledge, we propose the first self-stabilizing Byzantine-tolerant algorithm for solving MVC in asynchronous message-passing systems, enriched with required primitives.
	\emsD{That is, our} solution is built on using an SSBFT Binary consensus object, a BV-broadcast object, and two SSBFT BRB objects as well as a synchronous recycling mechanism. 
	\emsD{We believe that our solution can stimulate research for the design of algorithms that can recover after the occurrence of transient faults.}
	
	\smallskip
	\noindent
	\emsF{\extendedAbstract{Due to the page limit, some of the proof details appear in the complementary technical report~\cite{DBLP:journals/corr/abs-2110-08592}.}}
	\emsD{
		For the reader's convenience, all abbreviations are listed below. 
		\underline{Glossary:}
		\textbf{ACAF}: asynchronous cycles (\reduce{while }assuming fairness);  
		\textbf{BFT}: (non-stabilizing) Byzantine fault-tolerant;
		\textbf{BRB}: Byzantine-tolerant Reliable Broadcast;
		\textbf{BV-broadcast}:  Binary-values broadcast;
		\textbf{CRWF}: communication rounds (without\reduce{ assuming} fairness);
		\textbf{DRS}: SSBFT BRB by Duvignau, Raynal, and Schiller~\cite{DBLP:journals/corr/abs-2201-12880};
		\textbf{GMRS}: SSBFT MMR by Georgiou, Marcoullis, Raynal, and Schiller~\cite{DBLP:conf/netys/GeorgiouMRS21} for Binary consensus and BV-broadcast;
		\textbf{MR}: the studied solution by Most{\'{e}}faoui and Raynal~\cite{DBLP:conf/opodis/MostefaouiR10};
		\textbf{MVC}: multivalued consensus (\SectionAbv~\ref{sec:backgroundMotivation});  
		\textbf{SSBFT}: self-stabilizing Byzantine fault-tolerant;
		\textbf{VBB}: Validated Byzantine \emsF{Broadcast, \eg BFT and SSBFT ones in Algorithms~\ref{alg:vbbBroadcast} and~\ref{alg:SSBFTVBB}, \respectivelyP}}

	\Section{System Settings}
	\label{sec:sys} 
	We consider an asynchronous message-passing system that has no guarantees of communication delay. Also, \emsB{the algorithms do not} access the (local) clock (or use timeout mechanisms). The system consists of a set, $\sP$, of $n$ nodes (or processes) with unique identifiers. Any (ordered) pair of nodes $p_i,p_j \in \sP$ has access to a unidirectional communication channel, $\mathit{channel}_{j,i}$, that, at any time, has at most $\capacity \in \bZ^+$ packets on transit from $p_j$ to $p_i$ (this assumption is due to a known impossibility~\cite[Chapter 3.2]{DBLP:books/mit/Dolev2000}).

	%
	\label{sec:interModel}
	We use the \emph{interleaving model}~\cite{DBLP:books/mit/Dolev2000} for representing the asynchronous execution of the system. The node's program is a sequence of \emph{(atomic) steps}. Each step starts with an internal computation and finishes with a single communication operation, \ie a message $send$ or $receive$. The \emph{state}, $s_i$, of node $p_i \in \sP$ includes all of $p_i$'s variables and $\mathit{channel}_{j,i}$. The term \emph{system state} (or configuration) refers to the tuple $c = (s_1, s_2, \cdots,  s_n)$. We define an \emph{execution (or run)} $R={c[0],a[0],c[1],a[1],\ldots}$ as an alternating sequence of system states $c[x]$ and steps $a[x]$, such that each $c[x+1]$, except for the starting one, $c[0]$, is obtained from $c[x]$ by $a[x]$'s execution.

	
	\Subsection{The fault model and self-stabilization}
	
	\emsE{We specify the fault model and design criteria.}
	
	\Subsubsection{Arbitrary node failures.}
	\label{sec:arbitraryNodeFaults}
	Byzantine faults model any fault in a node including crashes, and arbitrary malicious behaviors. Here the adversary lets each node receive the arriving messages and calculate its state according to the algorithm. However, once a node (that is captured by the adversary) sends a message, the adversary can modify the message in any way, delay it for an arbitrarily long period or even remove it from the communication channel. The adversary can also send fake messages spontaneously. The adversary has the power to coordinate such actions without any limitation.
	For the sake of solvability~\cite{DBLP:journals/toplas/LamportSP82,DBLP:journals/jacm/PeaseSL80,DBLP:conf/podc/Toueg84}, we limit the number, $t$, of nodes that can be captured by the adversary, \ie $n \geq 3t + 1$. The set of non-faulty indices is denoted by $\Correct$ and called the set of correct nodes.
	
	\Subsubsection{Arbitrary transient-faults}
	\label{sec:arbitraryTransientFaults}
	We consider any temporary violation of the assumptions according to which the system was designed to operate. 
	We refer to these violations and deviations as \emph{arbitrary transient-faults} and assume that they can corrupt the system state arbitrarily (while keeping the program code intact). 
	The occurrence of a transient fault is rare. 
	Thus, we assume that the last arbitrary transient fault occurs before the system execution starts~\cite{DBLP:books/mit/Dolev2000}. 
	Also, it leaves the system to start in an arbitrary state.
	In other words, we assume arbitrary starting states at all correct nodes and the communication channels that lead to them. 
	\emsD{Moreover, transient faults do not occur during the system execution.}
	
	\Subsubsection{Dijkstra's self-stabilization}
	\label{sec:Dijkstra}
	\emsD{The \emph{legal execution} ($LE$) set refers to all executions in which the problem requirements hold.}
	%
	%
	A system is \emph{self-stabilizing} with respect to $LE$, when every execution $R$ of the algorithm reaches within a finite period a suffix $R_{legal} \in LE$ that is legal. Namely, Dijkstra~\cite{DBLP:journals/cacm/Dijkstra74} requires $\forall R:\exists R': R=R' \circ R_{legal} \land R_{legal} \in LE \land |R'| \in \bZ^+$, where the operator $\circ$ denotes that $R=R' \circ R''$ is the concatenation of $R'$ with $R''$.
	The part of the proof that shows the existence of $R'$ is called \emph{Convergence} (or recovery), and the part that shows $R_{legal} \in LE$ is called \emph{Closure}.

	

	\Subsubsection{Complexity measures \rdA{and execution fairness}}
	\label{sec:complex}
	\rdA{We say that execution fairness holds \emsF{among} processes if \emsF{the scheduler enables any correct} process infinitely \emsF{often, \ie the scheduler cannot (eventually) halt the execution of non-faulty processes.}}
	The time between the invocation of an operation (such as consensus or broadcast) and the occurrence of all required deliveries is called operation latency. As in MR, we show that the latency is finite without assuming execution fairness.
	The term stabilization time refers to the period in which the system recovers after the occurrence of the last transient fault. 
	When estimating the stabilization time, our analysis assumes that all correct nodes complete roundtrips infinitely often with all other correct nodes. 
	However, once the convergence period is over, no fairness assumption is needed.
	\emsD{Then, the} stabilization time is measured in terms of \emph{asynchronous cycles}, \emsD{which we define next.}  
	All self-stabilizing algorithms have a do forever loop since these systems cannot be quiescent due to a well-known impossibility~\cite[Chapter 2.3]{DBLP:books/mit/Dolev2000}.
	\emsD{Also, the study algorithms allow nodes to communicate with each other via broadcast operation.}
	Let $num_b$ be the maximum number of \emsD{(underlying)} broadcasts per iteration of the do forever loop.
	The first asynchronous cycle, $R'$, of execution $R=R'\circ R''$ is the shortest prefix of $R$ in which every correct node is able to start and complete at least a constant number, $num_b$, of round-trips with every correct node.
	The second asynchronous communication round of $R$ is the first round of the suffix $R''$, and so forth.

	
	\Subsection{Building Blocks}
	\label{sec:sabb}
	Following Raynal~\cite{DBLP:books/sp/Raynal18}, \Figure~\ref{fig:suit} illustrates a protocol suite for SSBFT state-machine replication using total order broadcast.
	This order can be defined by instances of MVC objects, which in turn, invoke SSBFT Binary consensus, BV-broadcast, and SSBFT recycling subroutine\reduce{ for consensus objects} (GMRS~\cite{DBLP:conf/netys/GeorgiouMRS21,DBLP:conf/sss/GeorgiouRS23}) as well as SSBFT BRB (DRS~\cite{DBLP:journals/corr/abs-2201-12880}).
	
	\Subsubsection{SSBFT Byzantine-tolerant Reliable Broadcast (BRB)}
	\label{sec:BRBext}
	The\reduce{ communication} abstraction of (single instance) BRB allows every node to invoke the $\mathsf{broadcast}(v):v\in V$ and $\mathsf{deliver}(k):p_k \in \sP$ operations.
	
	\begin{definition}
		\label{def:prbDef}
		\reduce{The operations} $\mathsf{broadcast}(v)$ and $\mathsf{deliver}(k)$ should satisfy:
		\begin{itemize}
			\item \textbf{BRB-validity.~~} Suppose a correct node BRB-delivers message $m$ from a correct $p_i$. Then, $p_i$ had BRB-broadcast $m$.
			\item \textbf{BRB-integrity.~~} No correct node BRB-delivers more than once.
			\item \textbf{BRB-no-duplicity.~~} No two correct nodes BRB-deliver different messages from $p_i$ (which might be faulty).
			\item \textbf{BRB-completion-1.~~} Suppose $p_i$ is a correct sender. All correct nodes BRB-deliver from $p_i$ eventually.
			\item \textbf{BRB-completion-2.~~} Suppose a correct node BRB-delivers a message from $p_i$ (which might be faulty). All correct nodes BRB-deliver $p_i$'s message eventually.
		\end{itemize}
	\end{definition}
	
	We assume the availability of an SSBFT BRB\reduce{ implementation}, such as the one in~\cite{DBLP:journals/corr/abs-2201-12880}, which stabilizes within $\bigO(1)$ asynchronous cycles. Such implementation lets $p_i \in \sP$ to use the operation $\mathsf{deliver}_i(k)$ for retrieving the current return value, $v$, of the BRB broadcast from $p_k \in \sP$. 
	Before the completion of the task of the $\mathsf{deliver}_i(k)$ operation, $v$'s value is $\bot$. 
	This way, whenever $\mathsf{deliver}_i(k)\neq \bot$, node $p_i$ knows that the task is completed and the returned value can be used.
	There are several BRB implementations~\cite{DBLP:conf/podc/AlhaddadDD0VXZ22,DBLP:conf/ccs/DasX021,DBLP:conf/srds/MaurerT14} that satisfy different requirements than the ones in Definition~\ref{def:prbDef}\reduce{, which is taken from the textbook~\cite{DBLP:books/sp/Raynal18}}.
	
	Note that existing non-self-stabilizing BFT BRB\reduce{ implementations}, \eg~\cite[Ch. 4]{DBLP:books/sp/Raynal18}, consider another kind of interface between BRB and its application. In that interface, BRB notifies the application via the raising of an event whenever a new message is ready to be BRB-delivered. However, in the context of self-stabilization, a single transient fault can corrupt the BRB object to encode in its internal state that the message was already BRB-delivered without ever BRB-delivering the message. The interface proposed in~\cite{DBLP:journals/corr/abs-2201-12880} addresses this challenge by allowing the application to repeatedly query the status of the SSBFT BRB object without changing its state.
	
	\emsE{We also assume that BRB objects have the interface function $\hasTerminated()$, which serves as a predicate indicating when the sender knows that all non-faulty nodes have successfully delivered the application message. The implementation of $\hasTerminated()$ is straightforward — it checks the condition in the if-statement on line 49 of Algorithm 4 in~\cite{DBLP:journals/corr/abs-2201-12880}. If the condition is met, it returns $\false$, otherwise, it returns $\true$.}
	
	\Subsubsection{SSBFT Binary-values Broadcast (BV)}
	\label{sec:sefBVbrodcast}
	This is an all-to-all broadcast operation of Binary values. This abstraction uses the operation, $\mathsf{bvBroadcast}(v)$, which is assumed to be invoked independently (\ie not necessarily simultaneously) by all the correct nodes, where \emsD{$v \in V$. 
		For the sake of a simpler presentation of our solutions, we prefer $V=\{\False,\True\}$ over the traditional $V= \{0,1\}$ presentation.}
	The set of values that are BV-delivered to node $p_i$ can be retrieved via the function $\binValues_i()$, which returns $\emptyset$ before the arrival of any $\mathsf{bvBroadcast}()$ by a correct node. We specify under which conditions values are added to $\binValues()$.
	
	\begin{itemize}
		
		\item \textbf{BV-validity.} Suppose $v \in \binValues_i()$ and $p_i$ is correct. It holds that $v$ has been BV-broadcast by a correct node.
		
		\item \textbf{BV-uniformity.} $v \in \binValues_i()$ and $p_i$ is correct. Eventually $\forall j \in \Correct: v \in \binValues_j()$.
		
		\item \textbf{BV-completion.} Eventually, $\forall i \in \Correct:\binValues_i() \neq \emptyset$\reduce{ holds}.
		
	\end{itemize}
	
	The above requirements imply that eventually $\exists s \subseteq \{\false,\true\}: s \neq \emptyset \land \forall i \in \Correct: \binValues_i()=s$ and the set $s$ does not include values that were BV-broadcast only by Byzantine nodes. 
	The SSBFT BV-broadcast solution in~\cite{DBLP:conf/netys/GeorgiouMRS21} stabilizes within $\bigO(1)$ asynchronous cycles. 
	This implementation allows the correct nodes to repeat a BV-broadcast using the same BV-broadcast object. 
	\emsD{As mentioned in \SectionAbv~\ref{sec:ourSSBFTvarMR}, this allows the proposed solution to overcome challenges related to the corruption of the state of the SSBFT Binary consensus object, more details in \SectionAbv~\ref{sec:mvcInv}.}
	
	
	\Subsubsection{SSBFT Binary Consensus}
	As mentioned, the studied solution reduces MVC to Binary consensus by enriching the system model with a BFT object that solves Binary consensus (Definition~\ref{def:consensus}).
	\begin{definition}
		\label{def:consensus}
		Every $p_i\in \sP$ has to propose a value $v_i \in V=\{\False,\True\}$ via an invocation of  $\mathsf{propose}_i(v_i)$. Let $\mathit{Alg}$ be a Binary Consensus \rdA{(BC)} algorithm. $\mathit{Alg}$ has to satisfy \emph{safety}, \ie BC-validity and BC-agreement, and \emph{liveness}, \ie BC-completion.
		\begin{itemize}
			\item \textbf{BC-validity.~~} 
			The value $v \in \{\False,\True\}$ decided by a correct node is a value proposed by a correct node.
			\item \textbf{BC-agreement.~~} Any two correct nodes that decide, do so with identical decided values.
			\item \textbf{BC-completion.~~} All correct nodes decide.
		\end{itemize}
	\end{definition}
	
	We assume the availability of SSBFT Binary consensus, such as the one from GMRS~\cite{DBLP:conf/netys/GeorgiouMRS21}, which stabilizes within $\bigO(1)$ asynchronous cycles. GMRS's solution might fail to decide with negligible probability. In this case, GMRS's solution returns the error symbol, $\blitza$, instead of a legitimate value from the set $\{\false,\true\}$.
	The proposed SSBFT MVC algorithm returns $\blitza$ whenever the SSBFT Binary consensus returns $\blitza$ (cf. line~\ref{ln:normalReturnAndConsisT} of Algorithm~\ref{alg:consensus}).
	
	\Subsubsection{The Recycling Mechanism and Recyclable Objects}
	\label{sec:recyclable}
	Just as MR, we do not focus on the management of consensus invocations since we assume the availability of a mechanism for eventually recycling all MVC objects that have completed their tasks. 
	GMRS~\cite{DBLP:conf/netys/GeorgiouMRS21,DBLP:conf/sss/GeorgiouRS23} implement such subroutine. 
	We review their subroutine and explain how they construct recyclable objects.
	
	\emsD{GMRS implements consensus objects using a storage of constant size allocated at program compilation time. 
		Since these objects can be instantiated an unbounded number of times, it is necessary to reuse the storage once a consensus is reached. 
		This should occur only after each correct node received the decided value via $\mathsf{result}()$.}
	
	\emsD{To facilitate this, GMRS assumes that the object has two meta-statuses: \emph{used} and \emph{unused}. 
		The \emph{unused} status indicates the availability of objects that were either never used or are no longer in current use.
		GMRS specifies that recyclable objects must implement an interface function called $\mathsf{wasDelivered}()$, which returns $1$ after the result delivery.
		Recycling is triggered by the recycling mechanism, which invokes $\mathsf{recycle}()$ at each non-faulty node, setting the meta-status of the corresponding consensus object to \emph{unused}.}
	
	\emsD{GMRS defines recyclable object construction as a task that requires eventual agreement on the value of $\mathsf{wasDelivered}()$. 
		In detail, if a non-faulty node $p_i$ reports delivery (\ie $\mathsf{wasDelivered}_i()=1$), then all non-faulty nodes will eventually report delivery as well.
		We clarify that during the recycling process, \ie when at least one non-faulty node invokes $\mathsf{recycle}()$, there is no need to maintain agreement on the values of $\mathsf{wasDelivered}()$.}
	
	\emsD{GMRS requires us to implement 
		$\mathsf{recycle}()$ by locally setting the algorithm to its predefined post-recycling state, see \SectionsAbv~\ref{sec:LocVarVBB} and~\ref{sec:LocVarMVC}. 
		Also, our solution implements\reduce{ the operation} $\done()$, which facilitates the implementation of $\mathsf{wasDelivered}()$ using the same construction proposed by GMRS in~\cite{DBLP:conf/netys/GeorgiouMRS21,DBLP:conf/sss/GeorgiouRS23}.
		By implementing GMRS' interfaces, we borrow GMRS correctness properties since the studied problem and the structure of the proposed algorithms are very similar.} 
	
	\begin{algorithm}[t!]
		\begin{\algSize}	
			
			\smallskip
			
			\textbf{operation} $\mathsf{vbbBroadcast}(v)$ \label{ln:vbbBradcastAAA} \Begin{
				
				BRB-broadcast $\mathrm{INIT}(i, v)$\label{ln:brbBradcast0}\; 
				
				\textbf{wait} $|rec|\geq n\mathit{-}t$ \textbf{where} $rec$ is the multiset of BRB-delivered values\label{ln:brbBradcast0wait}\;
				
				
				BRB-broadcast $\mathrm{VALID}(i, (\mathit{equal}(v, rec) \geq  n \mathit{-} 2t))$\label{ln:brbBradcast1}\; 
				
			}
			
			\smallskip
			
				
				\ForEach{$p_j \in \sP$ \emph{execute concurrently}\label{ln:vbbBackground}}{
					
					\textbf{wait} $\mathrm{INIT}(j,v)$ and $\mathrm{VALID}(j,x)$ BRB-delivered from $p_j$\label{ln:vbbWaitValidINIT}\;
					
					\If{$x$\label{ln:ifXtrue}}{\{\textbf{wait} $(\mathit{equal}(v, rec) \geq n \mathit{-} 2t)$; $d \gets v$\}} 
					\Else{\{\textbf{wait} $(\mathit{differ}(v, rec) \geq t \mathit{+} 1)$; $d \gets \blitza$\label{ln:ifXtrueElse}\}}
					
					$\mathsf{vbbDeliver}(d)$ at $p_i$ as the value VBB-broadcast by $p_j$\label{ln:vbbDeliverA}\;
					

			}
			
			
			\caption{\label{alg:vbbBroadcast}Non-self-stabilizing BFT VBB-broadcast; code for $p_i$}
		\end{\algSize}
	\end{algorithm}

	\emsD{GMRS implements a recycling service using\remove{ synchrony assumptions and} a synchronous SSBFT consensus that allows all non-faulty nodes to reuse the object immediately after the process returns from $\mathsf{recycle}()$.}
	GMRS's recycling facilitates the transformation of the non-self-stabilizing BFT MR algorithm to an SSBFT one.
	The transformation concentrates on assuring\reduce{ operation} completion since once all objects have been recycled, the system reaches its \emph{post-recycling state}, which has no \rd{trace} of stale information, \ie Convergence holds. 
	\emsD{As mentioned in \SectionAbv~\ref{sec:ourSSBFTvarMR}, the} effect of these assumptions can be mitigated by letting recycling batches of $\delta$ objects, where $\delta$ is a predefined constant that depends on the available memory. 
	This way, the communication-intensive components are asynchronous and the synchronous recycling actions occur according to a load that is defined by $\delta$.

	\Section{The Studied Algorithms}
	\label{sec:algos}
	As mentioned, MR is based on a reduction of BFT MVC to BFT Binary consensus. 
	MR guarantees that the decided value is not a value proposed only by Byzantine nodes. 
	Also, if there is a value, $v \in V$, that all correct nodes propose, then $v$ is decided. 
	Otherwise, the decided value is either a value proposed by the correct nodes or~\reduce{ the error symbol,} $\blitza$. 
	This way, an adversary that commands its captured nodes to propose the same value, say, $v_{byz} \in V$, cannot lead to the selection of $v_{byz}$ without the support of at least one correct node. 
	MR uses the VBB communication abstraction (\Figure~\ref{fig:suit}), which we present (\SectionAbv~\ref{sec:VBB}) before we bring the reduction algorithm (\SectionAbv~\ref{sec:MVC}).
	
	\Subsection{Validated Byzantine Broadcast (VBB)} 
	\label{sec:VBB}
	This abstraction sends messages from all nodes to all nodes. 
	It allows the operation, $\mathsf{vbbBroadcast}(v)$ and raises the event $\mathsf{vbbDeliver}(d)$, for VBB-broadcasting, and \respectivelyC VBB-delivering.

	\Subsubsection{\ems{Specifications}} 
	\label{sec:specVVB}
	We detail VBB-broadcast requirements.
	
	\begin{itemize}
		\item \textbf{VBB-validity.~~} VBB-delivery of messages needs to relate to VBB-broadcast of messages in the following manner.
		\begin{itemize}
			\item \textbf{VBB-justification.~~} Suppose $p_i : i \in \Correct$ VBB-delivers message $m\neq \blitza$ from some (faulty or correct) node. There is at least one correct node that VBB-broadcast $m$.
			
			\item \textbf{VBB-obligation.} Suppose all correct nodes VBB-broadcast\reduce{ the same} $v$. All correct nodes VBB-delivers $v$ from each correct node.
		\end{itemize}
		
		\item \textbf{VBB-uniformity.~~}  Let $p_i:i \in \Correct$. Suppose VBB-delivers $m' \in \{m,\blitza\}$ from a (possibly faulty)\reduce{ node} $p_j$. All the correct nodes VBB-deliver the same message $m'$ from $p_j$.
		
		\item \textbf{VBB-completion.~~} Suppose a correct node $p_i$ VBB-broadcasts $m$. All the correct nodes VBB-deliver from $p_i$.
	\end{itemize}
	
	We also say that a complete VBB-broadcast instance includes $\mathsf{vbbBroadcast}_i(m_i)$ invocation by every correct $p_i \in \sP$.
	It also includes $\mathsf{vbbDeliver}()$ of $m'$ from at least $(n\mathit{-}t)$ distinct nodes, where $m'$ is either $p_j$'s message, $m_j$, or~\reduce{ the error symbol,} $\blitza$. 
	The latter~\reduce{ value} is returned when a message from a given sender cannot be validated. 
	This validation requires $m_j$ to be VBB-broadcast by at least one correct node.

	\Subsubsection{Implementing VBB-broadcast}
	Algorithm~\ref{alg:vbbBroadcast} presents the studied VBB-broadcast. 
	
	
	\noindent \textbf{Notation:~~} 
	%
	\emsC{Denote by} $\mathit{equal}(v, rec)$ and $\mathit{differ}(v, rec)$ the number of items in multiset $rec$ that are equal to, and \respectivelyC different from $v$. 
	
	\noindent \textbf{Overview:~~} 
	Algorithm~\ref{alg:vbbBroadcast} invokes BRB-broadcast twice in the first part of the algorithm (lines~\ref{ln:vbbBradcastAAA} to~\ref{ln:brbBradcast1}) and then VBB-delivers messages from nodes in the second part (lines~\ref{ln:vbbBackground} to~\ref{ln:vbbDeliverA}).
	
	Node $p_i$ first BRB-broadcasts $\mathrm{INIT}(i, v_i)$ (where $v_i$ is the VBB-broadcast message), and suspends until the arrival of $\mathrm{INIT}()$ from at least $(n \mathit{-} t)$ different nodes (lines~\ref{ln:brbBradcast0} to~\ref{ln:brbBradcast0wait}), which $p_i$ collects in the multiset $rec_i$. In line~\ref{ln:brbBradcast0}, node $p_i$ tests whether $v_i$ was  BRB-delivered from at least $n\mathit{-}2t \geq t\mathit{+}1$ different nodes. Since this means that $v_i$ was BRB-broadcast by at least one correct node, $p_i$ attests to the validity of $v_i$ (line~\ref{ln:brbBradcast1}). Recall that each time $\mathrm{INIT}()$ arrives at $p_i$, the message is added to $rec_i$. Therefore, the fact that $|rec_i| \geq n \mathit{-} t$ holds (line~\ref{ln:brbBradcast0wait}) does not keep $rec_i$ from growing.
	
	Algorithm~\ref{alg:vbbBroadcast}'s second part (lines~\ref{ln:vbbBackground} to~\ref{ln:vbbDeliverA}) includes $n$ concurrent background tasks. Each task aims at VBB-delivering a message from a different node, say, $p_j$. It starts by waiting until $p_i$ BRB-delivered both $\mathrm{INIT}(j, v_j)$ and $\mathrm{VALID}(j, x_j)$ from $p_j$ so that $p_i$ has both $p_j$'s VBB's values, $v_j$, and the result of its validation test, $x_j$. 
	
	\begin{enumerate}
		\item \label{itm:xTrue}\textbf{The $x_j=\true$ case (line~\ref{ln:ifXtrue}).~~} Since $p_j$ might be faulty, we cannot be sure that $v_j$ was indeed validated. Thus, $p_i$ re-attests $v_j$ by waiting until $\mathit{equal}(v_j, rec_i) \geq n\mathit{-}2t$ holds. If this happens, $p_i$ VBB-delivers $v_j$ as a message from $p_j$, because this implies $\mathit{equal}(v_j, rec_i) \geq t \mathit{+} 1$ since $n\mathit{-}2t \geq t\mathit{+}1$.
		
		\item \label{itm:xFalse}\textbf{The $x_j=\false$ case (line~\ref{ln:ifXtrueElse}).~~} For similar reasons to the former case, $p_i$ waits until $rec_i$ has at least $t\mathit{+}1$ items that are not $v_j$. This implies at least one correct note cannot attest $v_j$'s validity. If this ever happens, $p_i$ VBB-delivers the error symbol, $\blitza$, as the received message from $p_j$.
	\end{enumerate}
	
	

	\begin{algorithm}[t!]
		\begin{\algSize}

			\smallskip
			
			\textbf{variables:}
			$\mathit{bcO}$; \texttt{//} Binary consensus object, $\bot$ is the initial state
			
			\smallskip
			
			\textbf{macro} $\sameValue()$ \textbf{do return} $\exists v \neq$ $\blitza:\mathit{equal}(v, rec)$ $\geq$ $ n \mathit{-} 2t \land$ $rec=\{v'$ $\neq$ $\blitza\}$\label{ln:mvcLet} \textbf{where} $rec$ is a multiset of the $(n\mathit{-}t)$ values VBB-delivered (line~\ref{ln:mvcWait});
			
			\smallskip
			
			\textbf{operation} $\mathsf{propose}(v)$ \label{ln:mvcPropuse} \Begin{
				
				$\mathsf{vbbBroadcast}$ $\mathrm{EST}(v)$\label{ln:mvcESTsend}\; 
				
				\textbf{wait} $\mathrm{EST}(\bullet)$\reduce{ messages} VBB-delivered from $(n\mathit{-}t)$ different nodes\label{ln:mvcWait}\;
				
				
				\If{$\neg bcO.\mathsf{propose}(\sameValue())$\label{ln:mvcIf}}{\Return{$\blitza$}\label{ln:mvcThen}}
				
				\Else{\textbf{wait} $(\exists v\neq \bot$$:\mathit{equal}(v,rec)\geq n \mathit{-}2t) \textbf{ then } $\Return{$v$}\label{ln:mvcElse}}
				
			}

			
			\caption{\label{alg:reductionNon}Non-self-stabilizing BFT MVC; code for $p_i$}
		\end{\algSize}
	\end{algorithm}
	
	\Subsection{Non-stabilizing BFT Multivalued Consensus}
	\label{sec:MVC}
	Algorithm~\ref{alg:reductionNon} reduces the BFT MVC problem to BFT Binary consensus in\reduce{ message-passing} systems that have up to $t < n/3$ Byzantine nodes.
	\emsA{Algorithm~\ref{alg:reductionNon} uses VBB-broadcast abstraction (Algorithm~\ref{alg:vbbBroadcast}).} 
	Note that the line numbers of Algorithm~\ref{alg:reductionNon} continue the ones of Algorithm~\ref{alg:vbbBroadcast}. 
	
	\Subsubsection{\ems{Specifications}} 
	\emsC{Our BFT MVC task (\SectionAbv~\ref{sec:backgroundMotivation})} includes the requirements of BC-validity, BC-agreement, and BC-completion (\SectionAbv~\ref{sec:backgroundMotivation}) as well as the BC-no-Intrusion property (\SectionAbv~\ref{sec:BYZnoIntro}). 
	
	\Subsubsection{\ems{Implementation}} 
	\label{sec:impNonMPC}
	%
	%
	%
	Node $p_i$ waits for $\mathrm{EST}()$ messages from $(n \mathit{-} t)$ different nodes after it as VBB-broadcast its own value (lines~\ref{ln:mvcESTsend} to~\ref{ln:mvcWait}). It holds all the VBB-delivered values in the multiset $rec_i$ (line~\ref{ln:mvcLet}) before testing whether $rec_i$ includes (1) non-$\blitza$ replies from  at least $(n \mathit{-} 2t)$\reduce{ different} nodes, and (2)\reduce{ exactly} one non-$\blitza$ value $v$ (line~\ref{ln:mvcLet}). The test result is proposed to the Binary consensus object, $\mathit{bcO}$ (line~\ref{ln:mvcIf}).
	
	Once consensus is reached, $p_i$ decides according to\reduce{ the consensus result,} $\mathit{bcO}_i.\bcdone()$. Specifically, if $\mathit{bcO}_i.\bcdone() =\false$, $p_i$ returns the error symbol, $\blitza$, since there is no guarantee that any correct node was able to attest to the validity of the proposed value. Otherwise, $p_i$ waits until it received $\mathrm{EST}(v)$ messages that have identical values from at least $(n \mathit{-} 2t)$ different nodes (line~\ref{ln:mvcElse}) before returning that value $v$. Note that some of these $(n \mathit{-} 2t)$ messages were already VBB-delivered at line~\ref{ln:mvcWait}. The proof in~\cite{DBLP:conf/opodis/MostefaouiR10} shows that any correct node that invokes $\mathit{bcO}_i.\mathsf{propose}(\true)$ does so if all correct nodes eventually VBB-deliver identical values at least $(n \mathit{-} 2t)$ times. Then, any correct node can decide on the returned value for the MVC object once it also VBB-delivers identical values at least $(n \mathit{-} 2t)$ times.

%


	\Section{SSBFT Multivalued Consensus}
	
	Algorithms~\ref{alg:SSBFTVBB} and~\ref{alg:consensus} present our SSBFT VBB solution and self-stabilizing Byzantine- and intrusion-tolerant solution for MVC.
	They are obtained from Algorithms~\ref{alg:vbbBroadcast} and~\ref{alg:reductionNon} via code transformation and the addition of necessary consistency tests (\SectionsAbv~\ref{sec:vbbInv} and~\ref{sec:mvcInv}).
	Note that the line numbers of \emsA{Algorithms~\ref{alg:SSBFTVBB} and~\ref{alg:consensus}} continue the ones of Algorithms~\ref{alg:reductionNon}, and \respectivelyC~\ref{alg:SSBFTVBB}.

	\remove{
		
		\Subsection{Demonstrating convergence in the studied architecture}
		
		\emsB{Thus, the correctness proof needs to demonstrate convergance} by showing that any operation eventually returns a value that indicates operation completion, \emsB{regardless of the starting system state, which can be corrupted.} In other words, we demonstrate that when starting in an arbitrary system state, eventually, all objects become recyclable. As explained above, by eventually recycling all of these objects, the system arrives at a post-recycling state. For the sake of completeness, our proof also shows that, starting at a post-recycling state, the system satisfies the task requirements, which is MVC.
		
		We clarify that we do not deviate from the analytical framework proposed by Arora and Gouda~\cite{DBLP:journals/tse/AroraG93}, which requires the demonstration of the Closure and the Convergence properties. \emsB{Specifically,} our correctness proof demonstrates Convergence by showing that \emsB{all operations complete their operations eventually. This implies that} the components used and proposed by our solution always eventually become recyclable. Once they are all recycled, the system is in its post-recycling state. Starting from that state, Closure is proved. 
		
	} 

	\Subsection{SSBFT VBB-broadcast}
	The operation $\mathsf{vbbBroadcast}(v)$ allows the invocation of a VBB-broadcast instance with the value $v$. Node $p_i$ VBB-delivers messages from $p_k$ via $\mathsf{vbbDeliver}_i(k)$.

	\Subsubsection{\emsA{Algorithm~\ref{alg:vbbBroadcast}'s invariants that transient faults can violate}}
	\label{sec:vbbInv}
	\emsB{Transient faults can violate the following invariants, which our SSBFT solution addresses via consistency tests.}
	
	\begin{enumerate}
		\item\label{itm:ValidWithoutInit} Node $p_i$'s state must not encode the occurrence of BRB execution of phase $\valid$ (line~\ref{ln:brbBradcast1}) without encoding BRB execution of phase $\init$ (line~\ref{ln:brbBradcast0}). 
		\emsB{Algorithm~\ref{alg:SSBFTVBB} addresses this concern by informing that the VBB object has an internal error (line~\ref{ln:conTestBRB}). This way, indefinite blocking of the application is avoided.}
		
		\item\label{itm:badFormat} 
		\emsB{Define the phase types, $\texttt{vbbMSG} :=\{\init, \valid\}$ (line~\ref{ln:types}) for BRB dissemination of $\mathrm{INIT}()$, and \respectivelyC $\mathrm{VALID}()$ messages in Algorithm~\ref{alg:vbbBroadcast}.}
		For a given phase, $\typ\in\texttt{vbbMSG}$, the BRB message format must follow the one of BRB-broadcast of phase $\typ$, \emsB{as in lines~\ref{ln:brbBradcast0} and~\ref{ln:brbBradcast1}.
			In order to avoid blocking, the VBB object informs about an internal error (lines~\ref{ln:ilegalInputtoBRB} and~\ref{ln:vNotBotBlitza}).}
		
		\item\label{itm:nextPhase} \emsA{For a given phase, $\typ\in\texttt{vbbMSG}$, if at least $n-t$ different nodes BRB-delivered messages\reduce{ of phase $\typ$,} to node $p_i$, $p_i$'s state must lead to the next phase, \ie from $\init$ to $\valid$, or from $\valid$ to operation complete, in which VBB-deliver a non-$\bot$ value.} 
		\emsB{Algorithm~\ref{alg:SSBFTVBB} addresses this concern by monitoring the conditions in which the nodes should move from phase $\init$ to $\valid$ (line~\ref{ln:brbValid}). The case in which the nodes should move from phase $\valid$ to operation complete is more challenging since a single transient fault can (undetectably) corrupt the state of the BRB objects. Algorithm~\ref{alg:SSBFTVBB} makes sure that such inconsistencies are detected eventually. When an inconsistency is discovered, the VBB object informs\reduce{ the application} about an internal error (line~\ref{ln:blitzaVBBecho}), see \SectionAbv~\ref{sec:complicatInco} for more details.}	
	\end{enumerate}

	%
	%
	%
	
	\begin{algorithm}[t!]
		\begin{\algSize}

			\smallskip
			
			\textbf{types:}
			\label{ln:types} $\texttt{vbbMSG} :=\{\init, \valid\}$;  \emsB{\texttt{//} BRB object phases}
			

			%
			%
			
			\smallskip
			
			\textbf{variables:}
			
			$\mathit{brb}[\texttt{vbbMSG}][\sP]$ \emsB{\texttt{//} $\mathit{brb}[\init][\sP]$ and $\mathit{brb}[\valid][\sP]$ are\reduce{ the two} BRB objects. Upon recycling, \rd{ $[[\bot,\ldots,\bot],[\bot,\ldots,\bot]]$ is assigned}}
			

			
			
		
		\smallskip
		
		\textbf{macros:}
		
		\emsB{\texttt{//} exit conditions of wait operations in lines~\ref{ln:brbBradcast0wait} and~\ref{ln:ifXtrueElse}} 
		
		$\mathit{vbbWait}(k,\typ) := \exists {S \subseteq\sP: n \mathit{-}t\leq |S|}: \forall p_\ell \in S : (\mathit{brb}[\typ][k].\mathsf{deliver}(\ell) \neq \bot)$\label{ln:auxVSsP2tpkS}
		
		
		\smallskip
		\emsB{\texttt{//} detailed version of equal condition in lines~\ref{ln:brbBradcast1} and~\ref{ln:ifXtrue}} 
		
		$\mathit{vbbEq}(k,\typ,v) := \exists S \subseteq\sP: n \mathit{-}2t\leq |S| : \forall p_\ell \in S : ((\bull,v) = \mathit{brb}[\typ][k].\mathsf{deliver}(\ell))$\label{ln:vbbEq}\; 
		
		\smallskip
		
		\emsB{\texttt{//} detailed version of the differ condition used in line~\ref{ln:ifXtrueElse}} 
		
		$\mathit{vbbDiff}(k,\typ,v) :=\exists {S \subseteq\sP: t\mathit{+}1\leq |S|} : \forall {p_\ell \in S} : (v \neq \mathit{brb}[\typ][k].\mathsf{deliver}(\ell))$;


		\smallskip

		\textbf{operation:}
		$\mathsf{vbbBroadcast}(v)$ \label{ln:bvBradcast} \textbf{do} 
			$\mathit{brb}[\init][i].\mathsf{broadcast}((i,v))$ \texttt{//} cf. line~\ref{ln:brbBradcast0}
		
		\smallskip
		
		\textbf{do-forever} \Begin{
			
			
			\If{$\mathit{vbbWait}(i,(i,\initE)) \land$ $\mathit{brb}[\initE][i].\hasTerminated()$\label{ln:brbValid}}{
				\textbf{let} $v :=\mathit{brb}[\valid][i].\mathsf{deliver}(i)$\;
				$\mathit{brb}[\valid][i].\mathsf{broadcast}((i, \mathit{vbbEq}(i,\init,v)))$ \emsB{\texttt{//}\reduce{ cf.} line~\ref{ln:brbBradcast1}}\label{ln:brbValidThen}}
			
		}
		
		\smallskip
		
		\textbf{operation:} $\mathsf{vbbDeliver}(k)$ \label{ln:vbbDeliver} \Begin{
			
			\texttt{//} case (I) of the consistency tests (\SectionAbv~\ref{sec:vbbInv})
			
			\lIf{$\mathit{brb}[\initE][k].\mathsf{deliver}(k)=\bot \land$ $\mathit{brb}[\validE][k].\mathsf{deliver}(k) \neq \bot$\label{ln:conTestBRB}}{\Return{$\blitza$}}

			\texttt{//} wait until $p_j$'s BRB objects have delivered, cf. line~\ref{ln:vbbWaitValidINIT}
			
			\lIf{$\mathit{brb}[\initE][k].\mathsf{deliver}(k)=\bot\lor$ $\mathit{brb}[\validE][k].\mathsf{deliver}(k)=\emsB{\bot}$\label{ln:notReadyToReturn}}{\Return{$\bot$}}
			
			
			\smallskip
			
			
			\texttt{//} lines~\ref{ln:ilegalInputtoBRB} and~\ref{ln:vNotBotBlitzaIf} are case (II) of consistency tests (\SectionAbv~\ref{sec:vbbInv})

			
			
			
			\lIf{$\exists \typ\in\emph{\texttt{vbbMSG}}:\mathit{brb}[\typ][k].\mathsf{deliver}(k)=(j,\bull)\land$ $j\neq k$\label{ln:ilegalInputtoBRB}}{\Return{$\blitza$}}
			
			
			
			
			%
			%
			%
			%
			%
			
			\lIf{\emsA{$\neg ((\mathit{brb}[\initE][k].\mathsf{deliver}(k)=(k,v) \land v\in V)\land (\mathit{brb}[\validE][k].\mathsf{deliver}(k)=(k,x)  \land$ $x\in \{\false,\true\}))$\label{ln:vNotBotBlitzaIf}}}{\Return{$\blitza$}\label{ln:vNotBotBlitza}}	
			\lElseIf(\emsB{\texttt{//} cf. line~\ref{ln:ifXtrue}}){$x \land \mathit{vbbEq}(k,\validE,v)$\label{ln:xVBBeqRetVIF}}{\Return{$v$}\label{ln:xVBBeqRetV}}	
			\lElseIf(\emsB{\texttt{//} cf. line~\ref{ln:ifXtrueElse}}){$\neg x \land \mathit{vbbDiff}(k,\validE,v)$\label{ln:blitzaVBBdeliverIf}}{\Return{$\blitza$\label{ln:blitzaVBBdeliver}}}
			
			\texttt{//} case (III) of the consistency tests (\SectionAbv~\ref{sec:vbbInv})		
			
			\lElseIf{$\mathit{vbbWait}(k,\validE)$}{\Return{$\blitza$\label{ln:blitzaVBBecho}}}
			\Return{$\bot$\label{ln:vbbDeliverElad}} 	
			\texttt{//} $\mathsf{vbbDeliver}(k)$ is incomplete  
			
		}
		
		\smallskip

		
		\caption{\label{alg:SSBFTVBB}SSBFT VBB-broadcast; code for $p_i$}
	\end{\algSize}
\end{algorithm}

\Subsubsection{Local variables}
\label{sec:LocVarVBB}
The array $\mathit{brb}\texttt{vbbMSG}[\texttt{vbbMSG}][\sP]$ holds BRB object, which disseminate VBB-broadcast messages, \ie $\mathit{brb}[\init][]$ and $\mathit{brb}[\valid][]$ store the $\mathrm{INIT}()$, and \respectivelyC $\mathrm{VALID}()$ messages in Algorithm~\ref{alg:vbbBroadcast}.
\emsE{The second dimension of the array $\mathit{brb}[][]$ allows us to implement one VBB object per node as this is needed for Algorithm~\ref{alg:consensus}. 
	Thus, after the recycling of these objects (\SectionAbv~\ref{sec:recyclable}) or before they ever become active,} \rd{each of the $2n$ BRB objects has the value $\bot$}.
For a given $p_i \in \sP$, $\mathit{brb}_i[\bull][i]$ becomes active via the invocation of $\mathit{brb}_i[\bull][i].\mathsf{broadcast}(v)$ (which also leads to $\mathit{brb}_i[\bull][i]\neq\bot$) or the arrival of BRB protocol messages, say, from $p_j$ (which leads to $\mathit{brb}_i[\bull][j]\neq\bot$). 
Once a BRB message is delivered from $p_\ell$ \emsE{(in the context of phase $phs \in \texttt{vbbMSG}$ and VBB broadcast from $p_k$),} a call to $\mathit{brb}_i[phs][k].\mathsf{deliver}(\ell)$ retrieves the delivered message.  

\Subsubsection{Macros}
\emsB{The macro $\mathit{vbbWait}(\emsE{k,} \typ)$ (line~\ref{ln:auxVSsP2tpkS}) serves at\reduce{ if-statement} conditions in lines~\ref{ln:blitzaVBBecho} and~\ref{ln:brbValid} when the proposed transformation represents the exit conditions of the wait operations in lines~\ref{ln:brbBradcast0wait} and~\ref{ln:ifXtrueElse}. Specifically, given a phase, $\typ$, it tests whether there is a set $S$ that includes at least $n \mathit{-}t$ different nodes from which there is a message that is ready to be BRB-delivered.
	The macros $\mathit{vbbEq}(\emsE{k,} \typ,v)$ and $\mathit{vbbDiff}(\emsE{k,} \typ,v)$ are detailed versions of the equal, and \respectivelyC differ conditions used in lines~\ref{ln:brbBradcast1} and~\ref{ln:ifXtrue}, \respectivelyC line~\ref{ln:ifXtrueElse}. They check whether the value $v$ equals to, \respectivelyC differs from at least $n-2t$, \respectivelyC $t+1$ received BRB messages of phase $\typ$.}

\Subsubsection{The $\mathsf{vbbBroadcast}()$ operation (line~\ref{ln:bvBradcast})}
As in line~\ref{ln:brbBradcast0} in Algorithm~\ref{alg:vbbBroadcast}, $\mathsf{vbbBroadcast}(v)$'s invocation (line~\ref{ln:bvBradcast}) leads to \emsE{the invocation of the BRB object} $\mathit{brb}[\init][i].\mathsf{broadcast}((i,v))$.
Algorithm~\ref{alg:consensus} uses line~\ref{ln:brbValid} for implementing the logic of  lines~\ref{ln:brbBradcast0wait} and~\ref{ln:brbBradcast1} in Algorithm~\ref{alg:vbbBroadcast} \emsA{as well as the consistency test of item~\ref{itm:nextPhase} in \SectionAbv~\ref{sec:vbbInv}; that case of moving from phase $\init$ to $\valid$.} In detail, the macro $\mathit{vbbWait}(\emsE{k,} \typ)$ returns $\true$ whenever the BRB object $\mathit{brb}[\typ][k]$ has a message to BRB-deliver from at least $n-t$ different nodes. Thus, $p_i$ can ``wait'' for BRB deliveries from at least $n-t$ distinct nodes by testing \emsE{$\mathit{vbbWait}_i(\emsE{i,}\init)\land \mathit{brb}_i[\init][i].\hasTerminated()$, where the second clause indicates that $\mathit{brb}_i[\init][i]$ has terminated (\SectionAbv~\ref{sec:BRBext}), and thus, Item~\ref{itm:ValidWithoutInit} in \SectionAbv~\ref{sec:vbbInv} is implemented correctly.} Also,\reduce{ the macro} $\mathit{vbbEq}()$ is a detailed implementation of\reduce{ the function} $\mathit{equal}()$ (Algorithm~\ref{alg:vbbBroadcast}).




\Subsubsection{The $\mathsf{vbbDeliver}()$ operation (lines~\ref{ln:vbbDeliver} and~\ref{ln:vbbDeliverElad})}
\label{sec:complicatInco}
This operation (lines~\ref{ln:vbbDeliver} to~\ref{ln:vbbDeliverElad}) is based on lines~\ref{ln:vbbBackground} and~\ref{ln:vbbDeliverA} in Algorithm~\ref{alg:vbbBroadcast} together with a few of consistency tests \emsA{(\SectionAbv~\ref{sec:vbbInv}).}

\emsE{Line~\ref{ln:conTestBRB} performs a consistency test that matches Item~\ref{itm:ValidWithoutInit} in \SectionAbv~\ref{sec:vbbInv}, \ie for a given sender $p_k \in \sP$, if $p_k$ had invoked $\mathit{brb}[\validE][k]$ before $\mathit{brb}[\initE][k]$'s termination, an error is indicated via the return of $\blitza$.
	Line~\ref{ln:notReadyToReturn} follows line~\ref{ln:vbbWaitValidINIT}'s logic} by testing whether this VBB object is ready to complete w.r.t. sender $p_k \in \sP$. It does so by checking the state of the two BRB objects in $\mathit{brb}[\bull][k]$ since they each need to deliver a non-$\bot$ value. In case any of them is not ready to complete, the operation returns $\bot$.


\emsB{The if-statements in lines~\ref{ln:ilegalInputtoBRB} and~\ref{ln:vNotBotBlitzaIf} return $\blitza$ when the delivered BRB message is ill-formatted. By that, they fit the consistency test of item~\ref{itm:badFormat} in \SectionAbv~\ref{sec:vbbInv}.} 

The if-statements in lines~\ref{ln:xVBBeqRetVIF} to~\ref{ln:blitzaVBBdeliverIf} implement the logic of lines~\ref{ln:ifXtrue} to~\ref{ln:ifXtrueElse} in Algorithm~\ref{alg:vbbBroadcast}. 
The logic of these lines is explained in items~\ref{itm:xTrue}, and \respectivelyC~\ref{itm:xFalse} in \SectionAbv~\ref{sec:specVVB}.
Similar to line~\ref{ln:ifXtrue} in Algorithm~\ref{alg:vbbBroadcast}, $x_i$ \emsB{(line~\ref{ln:vNotBotBlitzaIf}) is the value} that $p_i$ BRB-delivers from $p_k$ via the\reduce{ BRB} object $\mathit{brb}_i[\valid]$. 
As mentioned, the macro $\mathit{vbbDiff}()$ is a detailed implementation of $\mathit{differ}()$ used by Algorithm~\ref{alg:vbbBroadcast}.

\ems{The if-statement in line~\ref{ln:blitzaVBBecho} considers the case in which $x_i$ is corrupted. Thus, there is a need to return\reduce{ the error symbol,} $\blitza$. This happens when $p_i$ VBB-delivered $\mathrm{VALID}()$ messages from at least $n\mathit{-}t$ different nodes, but none of the if-statement conditions in lines~\ref{ln:conTestBRB} to~\ref{ln:blitzaVBBdeliverIf} hold.} \emsA{This fits the consistency test of item~\ref{itm:nextPhase} in \SectionAbv~\ref{sec:vbbInv}, which requires eventual completion in the presence of transient faults.}

\begin{algorithm}[t!]
	\begin{\algSize}

		\smallskip
		
		\textbf{variables:}
		
		$\mathit{bvO}:=\bot$; \texttt{//} Binary-values object. \emsC{Recycling assigns $\bot$}
		
		$\mathit{bcO}:=\bot$; \texttt{//} Binary consensus object. \emsB{Upon recycling, assign $\bot$}
		
		\smallskip
		
		\textbf{macros:}
		
		\emsB{\texttt{//} exit conditions of the wait operation in line~\ref{ln:mvcWait}} 
		$\mathit{mcWait}() :=\exists S \subseteq\sP:n \mathit{-}t\leq |S| : \forall {p_k \in S} : (\mathsf{vbbDeliver}(k)\neq \bot)$\label{ln:mcEchoExists}\;
		
		\smallskip
		
		\texttt{//} adapted version of the same macro in line~\ref{ln:mvcLet} 
		
		$\sameValue()$ \label{ln:pbDef} \textbf{do} {\textbf{return} $(\exists v \notin \{\bot,\blitza\}: \exists S'\subseteq\sP:n \mathit{-}2t\leq $ $|S'|:\forall p_{k'} \in S' :$ $ (\mathsf{vbbDeliver}(k')=v)) \land $ $(|\{ \mathsf{vbbDeliver}(k) \notin \{\bot,\blitza\}:p_k \in \sP \}|=1)$\;
		}

		\smallskip
		
		\textbf{operation:}  			
		$\mathsf{propose}(v)$ \label{ln:mvcPropuseV}\textbf{do} $\mathsf{vbbBroadcast}(v)$;
		
		\smallskip
		
		\textbf{operation:}  $\done()$ \label{ln:mvcDone}\Begin{
			
			\texttt{//} test whether $\done()$ is not ready to complete 
			
			\If{$\mathit{bcO} = \bot \lor \mathit{bcO}.\bcdone() = \bot$\label{ln:notReady}}{\Return{$\bot$}}
			
			
			\ElseIf(\texttt{//} cf. line~\ref{ln:mvcThen}){$\neg \mathit{bcO}.\bcdone()$\label{ln:normalReturnAndConsis}}{\Return{$\blitza$}}
			
			
			\ElseIf(\texttt{//} cf. line~\ref{ln:mvcElse}){$\exists {v \notin \{\bot,\blitza\}} : \exists {S' \subseteq\sP: n \mathit{-}2t\leq |S'|} : \forall {p_{k'} \in S'} :  (\mathsf{vbbDeliver}(k')=v)$\label{ln:defultReturnIf}}{\label{ln:defultReturn}\Return{$v$}} 
			
			\smallskip
			
			\emsB{\texttt{//} perform a consistency test, cf. \SectionAbv~\ref{sec:mvcInv}}
			
			\ElseIf{$\mathit{mcWait}() \emsE{\land} \True \notin\mathit{bvO}.\binValues()$\label{ln:normalReturnAndConsisT}}{\Return{$\blitza$}}
			
			\Return{$\bot$;\label{ln:mvcDoneElad}} \emsB{\texttt{//} $\done()$ is not ready to complete}
			
		}		
		
		\smallskip
		
		\textbf{do-forever} \Begin{
			
			\If(\emsB{\texttt{//} cf. line~\ref{ln:mvcWait}}){$\mathit{mcWait}()$\label{ln:mcEcho}} {
				
				\lIf(\emsB{\texttt{//} cf. line~\ref{ln:mvcIf}}){$\mathit{bcO}= \bot$\label{ln:bcOpropuse}} {$\mathit{bcO}.\mathsf{propose}(\sameValue())$}
				
				$\mathit{bvO}.\mathsf{broadcast}(\sameValue())$\label{ln:bcOpropuseTest};  \emsB{\texttt{//} assist with\reduce{ the consistency test in} line~\ref{ln:normalReturnAndConsisT}}
				
			}
			
		}
		
		
		\caption{\label{alg:consensus}Self-stabilizing Byzantine-tolerant multivalued consensus via VBB-broadcast; code for $p_i$}
	\end{\algSize}
	
\end{algorithm}

\Subsection{SSBFT multivalued consensus}
The invocation of $\mathsf{propose}(v)$ VBB-broadcasts $v$. 
Node $p_i$ VBB-delivers messages from $p_k$ via the $\done_i()$ operation. 

\Subsubsection{\emsA{Algorithm~\ref{alg:reductionNon}'s invariants that transient faults can violate}}
\label{sec:mvcInv}
\emsB{As mentioned in \SectionAbv~\ref{sec:ourSSBFTvarMR}, the} occurrence of a transient fault can let the Binary consensus object to encode a decided value that was never proposed, \ie this violates BC-validity. 

Any SSBFT solution needs to address this concern since the MVC object can block indefinitely if $bcO$ decides $\True$ when $\forall p_j: j \in \Correct:\sameValue_j()=\False$ holds.
\emsB{As we explain next, our implementation BV-broadcasts (line~\ref{ln:bcOpropuseTest}) for testing the consistency of the SSBFT Binary consensus object (line~\ref{ln:normalReturnAndConsisT}). This way, indefinite blocking can be avoided by reporting an internal error state.}

%

\Subsubsection{Local variables}
\label{sec:LocVarMVC}
Algorithm~\ref{alg:consensus}'s state includes the SSBFT BV-broadcast object, $\mathit{bvO}$, and SSBFT consensus Binary object, $\mathit{bcO}$. 
Each has the post-recycling value of $\bot$\reduce{, \ie when $\mathit{bvO}=\bot$ (or $\mathit{bcO}=\bot$) the object is said to be inactive}.
They become active upon invocation and complete according to their specifications{ (\SectionsAbv~\ref{sec:BRBext} and~\ref{sec:sefBVbrodcast}, \respectively)}.

\Subsubsection{Macros}
\label{sec:MacMVC}
The macro $\mathit{mcWait}()$ (line~\ref{ln:mcEchoExists}) serves at the\reduce{ if-statement} conditions in lines~\ref{ln:normalReturnAndConsisT} and~\ref{ln:mcEcho} when the proposed transformation represents the exit conditions of the wait operations in lines~\ref{ln:mvcWait} and~\ref{ln:mvcElse}. Specifically, it tests whether there is a set $S \subseteq \sP$ that includes at least $n \mathit{-}t$ different nodes from which there is a message that is ready to be VBB-delivered.
The macro $\sameValue()$ is an adaptation of the macro in line~\ref{ln:mvcLet}, which tests whether there is a value $v \notin \{\bot,\blitza\}$ that a set of at least $n \mathit{-}2t$ different nodes have VBB-delivered and there is only one value $v' \notin \{\bot,\blitza\}$ that was VBB-delivered.

\Subsubsection{Implementation}
The logic of lines~\ref{ln:mvcPropuse} and~\ref{ln:mvcElse} in Algorithm~\ref{alg:reductionNon} is implemented by lines~\ref{ln:mvcPropuseV} to~\ref{ln:bcOpropuse} in Algorithm~\ref{alg:consensus}.
%
%
\Ie the invocation of $\mathsf{propose}(v)$ (line~\ref{ln:mvcPropuseV}) leads to the VBB-broadcast of $v$.

The logic of lines~\ref{ln:mvcWait} and~\ref{ln:mvcIf} in Algorithm~\ref{alg:reductionNon} is implemented by \emsB{lines~\ref{ln:mcEcho} and~\ref{ln:bcOpropuse}, \respectivelyP
	In detail, recall from \SectionAbv~\ref{sec:MacMVC} that $\mathit{mcWait}()$ (line~\ref{ln:mcEcho}) allows waiting until there are at least $n-t$ different nodes from which $p_i$ is ready to VBB-deliver a message.
	Then, if $\mathit{bcO}=\bot$ (\ie the Binary consensus object was not invoked), line~\ref{ln:bcOpropuse} uses $\mathit{bcO}$ to propose the returned value from $\sameValue()$. 
	Recall from \SectionAbv~\ref{sec:MacMVC},} the macro $\sameValue()$ (line~\ref{ln:pbDef}) implements the \emsB{one in line~\ref{ln:mvcLet} (Algorithm~\ref{alg:reductionNon}), see \SectionAbv~\ref{sec:impNonMPC} for details.}

Line~\ref{ln:bcOpropuseTest} facilitates the implementation of the consistency test (\SectionAbv~\ref{sec:mvcInv}) by using BV-broadcasting for disseminating the returned value from $\sameValue()$. 
This way it is possible to detect the case in which all correct nodes BV-broadcast a value that is, due to a transient fault, different than $\mathit{bcO}$'s decided one. 
\emsC{This is explained} when we discuss line~\ref{ln:normalReturnAndConsisT}.

The operation $\done()$ (lines~\ref{ln:mvcDone} to~\ref{ln:mvcDoneElad}) returns the decided value, which lines~\ref{ln:mvcThen} and~\ref{ln:mvcElse} implement in Algorithm~\ref{alg:reductionNon}. 
\emsB{It is a query-based operation, just as $\mathsf{deliver}()$ (cf. \emsC{text} just after Definition~\ref{def:prbDef}). 
	Thus,} line~\ref{ln:notReady} considers the case in which the decision has yet to occur, \ie it returns $\bot$. 
Line~\ref{ln:normalReturnAndConsis} considers the case that line~\ref{ln:mvcIf} (Algorithm~\ref{alg:reductionNon}) deals with and returns the error symbol, $\blitza$. 
Line~\ref{ln:defultReturn} implements line~\ref{ln:mvcElse} (Algorithm~\ref{alg:reductionNon}), \emsB{\ie it returns the decided value.}
\emsB{Line~\ref{ln:normalReturnAndConsisT} performs a consistency test for the case in which the \reduce{if-statment }conditions in lines~\ref{ln:notReady} to~\ref{ln:defultReturnIf} hold, there are VBB-deliveries from at least $n-t$ different nodes (\ie $\mathit{mcWait}_i()$ holds), and yet there are no correct node, say $p_j$, reports to $p_i$, via BV-broadcast, that the predicate $\sameValue_j()$ holds. 
	\technicalReport{Lemma~\ref{thm:mainMVCTerm} shows that this test addresses the challenge described in \SectionAbv~\ref{sec:mvcInv}.}}
\emsB{Whenever none of the conditions of the if-statements in lines~\ref{ln:notReady} to~\ref{ln:normalReturnAndConsisT} hold, line~\ref{ln:mvcDoneElad} returns $\bot$.}



\Section{Correctness}
%
%
%
%
%
%
\emsB{As explained in \SectionAbv~\ref{sec:recyclable}, we demonstrate Convergence (Theorem~\ref{thm:vbbTerminate}) by showing that all operations eventually complete since this implies their recyclability, and thus, the SSBFT object recycler can restart their state (\SectionAbv~\ref{sec:recyclable}).}
\extendedAbstract{For every layer, \ie VBB-broadcast and MVC, \emsF{there is a need to prove} the properties of completion and Convergence before demonstrating the Closure property. Due to the page limit, the proof of the Closure properties and the MVC's Convergence properties appear in the complementary technical report~\cite{DBLP:journals/corr/abs-2110-08592}.}
\technicalReport{For every layer, \ie VBB-broadcast and MVC, we \emsE{prove} the properties of completion and Convergence  (Theorems~\ref{thm:vbbTerminate} and \respectivelyC~\ref{thm:mvcTerminate}) before demonstrating the Closure property, cf. Theorems~\ref{thm:vbbClousre} and~\ref{thm:mvcClousre} \respectively}





\remove{
	\begin{Lemma}[BRB-no-duplicity]
		\label{thm:brbDuplicity}
		BRB-no-duplicity holds.
	\end{Lemma}
	\renewcommand{\lemcnt}{\ref{thm:brbDuplicity}}
	\begin{lemmaProof}
		Suppose that two correct nodes, $p_i$ and $p_j$, BRB-deliver $\langle j,m \rangle$ and $\langle j,m' \rangle$, \respectivelyP The proof needs to show that $m = m'$.
		If $p_i$ BRB-delivers $\langle j,m \rangle$, it received $\mathrm{MSG}(\mathit{mJ}):\mathit{mJ}[\typ][\ready]\supseteq \{(j,m)\}$ from $(2t\mathit{+}1)$ different processes, and hence from at least one non-faulty process. Similarly, $p_j$ received $\mathrm{MSG}(\mathit{mJ}):\mathit{mJ}[\typ][\ready]\supseteq \{(j,m')\}$ from at least one non-faulty process. It follows from Claim~\ref{thm:brbDuplicity} that all correct nodes broadcast $\mathrm{MSG}(\mathit{mJ}):\mathit{mJ}[\typ][\ready]\supseteq \{(j,v)\}$, from which we conclude that $m = v$ and $m' = v$.
	\end{lemmaProof}
} 

\Subsection{VBB-completion and Convergence}
\label{sec:CompletionVBB}
The proof demonstrates Convergence by considering executions that start in arbitrary states.
Theorem~\ref{thm:vbbTerminate} shows that all VBB objects are completed within a bounded time. 
Specifically, assuming fair execution among the correct nodes (\SectionAbv~\ref{sec:complex}), Theorem~\ref{thm:vbbTerminate} shows that, within a bounded time, for any pair of correct nodes, $p_i$, and $p_j$, a non-$\bot$ value is returned from $\mathsf{vbbDeliver}_j(i)$. 
As explained in \SectionAbv~\ref{sec:recyclable}, this means that all VBB objects become recyclable, \ie $\mathsf{wasDelivered}_i()$ returns $\true$. 
Since we assume the availability of the object recycling mechanism, the system reaches a post-recycling state within a bounded time.
Specifically, using the mechanism by GMRS~\cite{DBLP:conf/netys/GeorgiouMRS21,DBLP:conf/sss/GeorgiouRS23}, Convergence is completed with $\bigO(t)$ synchronous rounds.
We introduce the \emph{CRWF/ACAF} notation since the arguments of Theorem~\ref{thm:vbbTerminate} can be used for demonstrating different properties under different assumptions. Specifically, Theorem~\ref{thm:vbbTerminate} demonstrates that VBB-completion occurs within $\bigO(1)$ communication rounds (\SectionAbv~\ref{sec:complex}) without assuming execution fairness but assuming that execution $R$ starts in a post-recycling system state.
For the sake of brevity, when the proof arguments are used for counting the number of Communication Rounds Without assuming Fairness (CRWF), we write `within $\bigO(1)$ CRWF'.
Theorem~\ref{thm:vbbTerminate} also demonstrates Convergence within $\bigO(1)$ asynchronous cycles assuming fair execution among the correct nodes (\SectionAbv~\ref{sec:complex}). Thus, when the proof arguments can be used for counting the number of Asynchronous Cycles while Assuming Fairness (ACAF), we say, in short, `within $\bigO(1)$ ACAF'.
Moreover, when the same arguments can be used in both cases, we say 'within $\bigO(1)$ CRWF/ACAF'.

\begin{Theorem}[VBB-completion]
	\label{thm:vbbTerminate}
	%
	%
	Suppose all correct nodes invoke $\mathsf{vbbBroadcast}()$ within $\bigO(1)$ CRWF/ACAF. 
	Then, $\forall_{i,j \in \Correct} : \mathsf{vbbDeliver}_j(i) \neq \bot$ holds within $\bigO(1)$ CRWF/ACAF.
\end{Theorem}	
\renewcommand{\thmcnt}{\ref{thm:vbbTerminate}}
\begin{theoremProof}
	Let $ i \in \Correct$.
	Suppose either $p_i$ VBB-broadcasts $m$ in $R$ or $\exists \typ \in \texttt{vbbMSG}: \mathit{brb}_j[\typ][i] \neq \bot$ holds in $R$'s starting state. 
	We demonstrate that all correct nodes VBB-deliver $m' \neq \bot$ from $p_i$ by considering all the if-statements in lines~\ref{ln:conTestBRB} to~\ref{ln:blitzaVBBecho} and showing that, within $\bigO(1)$ CRWF/ACAF, one of the if-statements \emsE{(that returns a non-$\bot$)} in lines~\ref{ln:conTestBRB} to~\ref{ln:blitzaVBBecho} holds.

	\smallskip
	\noindent \noindent \textbf{Argument 1.} \emph{Suppose \emsE{$\exists \typ \in \texttt{vbbMSG}: \exists \ell \in \Correct:\mathit{brb}_j[\typ][i]$ $ \neq \bot$, \emsE{\ie $\mathit{brb}_j[\typ][i]$ is not in its post-recycling state.}  
			Within $\bigO(1)$ CRWF/ACAF, $\forall {k \in \Correct} : \mathit{brb}_k[\typ][\ell] \neq \bot$} holds.~~}
	The proof is implied by BRB-completion-1, BRB-completion-2, and the $\bigO(1)$ stabilization time of SSBFT BRB, cf. \SectionAbv~\ref{sec:BRBext}.
	
	\smallskip
	\noindent \textbf{Argument 2.} \emph{Suppose in $R$, the\reduce{ if-statement} condition in line~\ref{ln:conTestBRB} does not hold. Within $\bigO(1)$ CRWF/ACAF,  $\mathit{brb}_i[\validE][i] \neq \bot$ holds.~~}
	By the theorem assumption that all correct nodes invoke $\mathsf{vbbBroadcast}()$ within $\bigO(1)$ CRWF/ACAF, the BRB properties (Definition~\ref{def:prbDef}), and that there are at least $(n \mathit{-} t)$ correct nodes, the\reduce{ if-statement} condition in line~\ref{ln:brbValid} holds within $\bigO(1)$ CRWF/ACAF.
	Then, $p_i$ invokes the operation $\mathit{brb}_i[\valid][i]\mathsf{broadcast}(\bull)$. 
	
	\smallskip
	\noindent \textbf{Argument 3.} \emph{Suppose the condition $\mathit{brb}_i[\validE][i] \neq \bot$ holds in $R$'s starting state. Within $\bigO(1)$ CRWF/ACAF, either the if-statement condition in line~\ref{ln:conTestBRB} holds or the one in line~\ref{ln:notReadyToReturn} cannot hold.~~}
	The proof is implied by Algorithm~\ref{alg:consensus}'s code, \emsE{BRB-completion, and arguments 1 and 2.}
	
	\smallskip
	\noindent \textbf{Argument 4.} \emph{Within $\bigO(1)$ CRWF/ACAF, $\mathsf{vbbDeliver}_j(i) \neq \bot$ holds.~~}
	
	\noindent
	Suppose the\reduce{ if-statement} conditions in lines~\ref{ln:conTestBRB} to~\ref{ln:blitzaVBBdeliver} never hold. 
	By $\mathit{vbbWait}()$'s definition (line~\ref{ln:auxVSsP2tpkS}), BRB properties (Definition~\ref{def:prbDef}), the presence of at least $n\mathit{-}t$ correct and active nodes, and arguments (1) to (3), the\reduce{ if-statement} condition in line~\ref{ln:blitzaVBBecho} holds within $\bigO(1)$ CRWF/ACAF.
\end{theoremProof}

\technicalReport{
	\Subsection{Closure of VBB-broadcast}
	\label{sec:VBBclosure}
	Theorem~\ref{thm:BCagreement} demonstrates Closure by considering executions that start from a post-recycling state, which Theorem~\ref{thm:vbbTerminate} implies that the system reaches, see \SectionAbv~\ref{sec:CompletionVBB} for details.
	Theorem~\ref{thm:vbbClousre}'s proof shows no consistency tests causes false error indications.
	Theorem~\ref{thm:vbbClousre} counts communication rounds (without assuming fairness) using the CRWF notation presented in \SectionAbv~\ref{sec:CompletionVBB}.
	
	\begin{Theorem}[VBB-Closure]  
		\label{thm:vbbClousre}
		Let $R$ be an Algorithm~\ref{alg:consensus}'s execution in which all correct nodes invoke $\mathsf{vbbBroadcast}()$ within $\bigO(1)$ CRWF. Assume $R$ starts in a post-recycling state. 
		%
		%
		$R$ satisfies the VBB requirements (\SectionAbv~\ref{sec:specVVB}).
	\end{Theorem}
	
	\renewcommand{\thmcnt}{\ref{thm:vbbClousre}}
	\begin{theoremProof}
		VBB-completion holds (Theorem~\ref{thm:vbbTerminate}).
		
		\begin{Lemma}[VBB-uniformity]
			\label{thm:vbbUniformity}
			VBB-uniformity holds. 
		\end{Lemma}	
		\renewcommand{\lemcnt}{\ref{thm:vbbUniformity}}
		\begin{lemmaProof}
			Let $ i \in \Correct$. 
			Suppose $p_i$ VBB-delivers $m' \in \{m,\blitza\}$ from a (possibly faulty) $p_j \in \sP$. 
			We show: all the correct nodes VBB-deliver the same message $m'$ from $p_j$.
			Since $R$ is post-recycling and $p_i$ VBB-delivers $m'$ from $p_j$, the condition \emsE{$\mathit{brb}_i[\init][j].\mathsf{deliver}(j)=\bot \land \mathit{brb}_i[\valid][j].\mathsf{deliver}(j) \neq \bot$} (\reduce{of the if-statement in }line~\ref{ln:conTestBRB}) cannot hold. And, within $\bigO(1)$ CRWF \emsE{we know that} $(\mathit{brb}_i[\init][j].\mathsf{deliver}()=(j,v_{j,i}) \land \mathit{brb}_i[\valid][j].\mathsf{deliver}()$ $=(j,x_{j,i}))$ (line~\ref{ln:vNotBotBlitzaIf}) hold (BRB-completion-2 and since all correct nodes invoke $\mathsf{vbbBroadcast}()$ within $\bigO(1)$ CRWF). 
			Also, \emsE{the condition in line~\ref{ln:conTestBRB} cannot hold \wrt $p_k$.} 
			And, within $\bigO(1)$ CRWF, $(\mathit{brb}_k[\init][j].\mathsf{deliver}()=(j,v_{j,k}) \land \mathit{brb}_k[\valid][j].\mathsf{deliver}()$ $=(j,x_{j,k}))$ holds, such that $v_{j,i}=v_{j,k}$ and $x_{j,i}=x_{j,k}$.
			This is because $R$ starts in a post-recycling system state, BRB-no-duplicity, and BRB-completion-2, which means that every correct\reduce{ node} $p_k$ BRB-delivers, within $\bigO(1)$ CRWF, the same messages that $p_i$ delivers. 
			Due to similar reasons, depending on the value of $x_{j,i}=x_{j,k}$, \emsE{one of} the conditions\reduce{ of the if-statements} in lines~\ref{ln:xVBBeqRetV} or~\ref{ln:blitzaVBBdeliver} must hold. \Ie $p_k$ VBB-delivers, within $\bigO(1)$ CRWF, the same value as $p_i$ does.
		\end{lemmaProof}
		
		\begin{Lemma}[VBB-obligation]
			\label{thm:vbbObligation}
			VBB-obligation holds.
		\end{Lemma} 
		\renewcommand{\lemcnt}{\ref{thm:vbbObligation}}
		\begin{lemmaProof}
			Suppose all correct nodes, $p_j$, VBB-broadcast the same value $v$. 
			We show: every correct node, $p_i$, VBB-delivers $v$ from $p_j$. 
			Since every correct node invokes $\mathsf{vbbBroadcast}(v)$ within $\bigO(1)$ CRWF, $p_j$ invokes $\mathit{brb}_j[\init][j].\mathsf{broadcast}((j,v))$ (line~\ref{ln:bvBradcast}). 
			Since $\exists_{S \subseteq\sP} n \mathit{-}t\leq |S|: \forall {p_k \in \sP}: \mathit{brb}_i[\init][j].\mathsf{deliver}(\emsE{k})\neq \bot:p_i \in \Correct$ holds due BRB-completion-1, the\reduce{ if-statement} condition in line~\ref{ln:brbValid} holds within $\bigO(1)$ CRWF. 
			And the multi-set $\{\mathit{brb}_j[\init][j].\mathsf{deliver}(\emsE{k})\}_{p_k \in \sP}$ has at least $(n\mathit{-}2t)$ appearances of $(\bull,v)$. Thus, $p_j$ BRB-broadcasts\reduce{ the message} $(\valid,(j,\true))$ (line~\ref{ln:brbValidThen}). And, for any pair $k,\ell \in \Correct$, $\mathit{brb}_k[\valid][j].\mathsf{deliver}(\ell)=(j,\true)$ holds within $\bigO(1)$ CRWF (due to BRB-validity and BRB-completion-1). Thus, within $\bigO(1)$ CRWF, none of the\reduce{ if-statement} conditions at lines~\ref{ln:conTestBRB} to~\ref{ln:vNotBotBlitzaIf} hold. 
			However, the one in line~\ref{ln:xVBBeqRetV} holds, within $\bigO(1)$ CRWF, and only for the value $v$. 
			Then, \emsE{$p_i$ VBB-delivers $v$ as a VBB-broadcast from $p_j$.}
		\end{lemmaProof}
		
		\begin{Lemma}[VBB-justification]
			\label{thm:vbbJustification}
			VBB-justification holds.
		\end{Lemma} 
		
		\renewcommand{\lemcnt}{\ref{thm:vbbJustification}}
		\begin{lemmaProof}
			Let $i \in \Correct$. Suppose $p_i$ VBB-delivers $m \notin \{\bot,\blitza\}$ in $a_i \in R$. 
			We show: a correct $p_j$ invokes $\mathsf{vbbBroadcast}_j(v):m=(j,v)$ in $a_j \in R$, such that $a_j$ appears in $R$ $\bigO(1)$ CRWF before $a_i$. 
			Since $m \notin \{\bot,\blitza\}$, $(\mathit{brb}_i[\init][j].\mathsf{deliver}(j)=(j,v) \land \mathit{brb}_i[\valid][j].\mathsf{deliver}(j)=(j,x))$ (line~\ref{ln:vNotBotBlitzaIf}) and $x \land \mathit{vbbEq}_i(j,\valid,$ $v)$ (line~\ref{ln:xVBBeqRetV}) hold, because only line~\ref{ln:xVBBeqRetV} returns (in $\mathsf{vbbDeliver}()$) neither $\bot$ nor $\blitza$ and it can only do so when the if-statement condition in line~\ref{ln:vNotBotBlitzaIf} does not hold. 
			Since $\mathit{vbbEq}_i(j,\valid,v)$ holds and $n \mathit{-} 2t \geq t \mathit{+} 1$, at least one correct node, say, $p_j$ had BRB-broadcast $v$ (both for the $\init$ and $\valid$ phases in lines~\ref{ln:bvBradcast}, and \respectivelyC~\ref{ln:brbValidThen}), because $R$ starts in a post-recycling state and by Theorem~\ref{thm:vbbTerminate}'s Argument (2). 
			Thus, $a_j$ is before $a_i$.
		\end{lemmaProof}
	\end{theoremProof}
	
	\Subsection{Convergence and completion of MVC}
	As in \SectionAbv~\ref{sec:CompletionVBB}, Convergence is proven by showing BC-completion in executions that start in any state.
	Due to the\reduce{ availability of the} recycling mechanism, once the task is completed, the system reaches a post-recycling state \emsD{(by GMRS's mechanism~\cite{DBLP:conf/netys/GeorgiouMRS21,DBLP:conf/sss/GeorgiouRS23}, which Convergences within $\bigO(t)$ synchronous rounds).}
	Theorem~\ref{thm:mvcTerminate} counts communication rounds \emsD{(without assuming fairness) and asynchronous cycles (while assuming fairness)} using the CRWF/ACAF notation (\SectionAbv~\ref{sec:CompletionVBB}.)
	
	\begin{Theorem}[BC-completion]
		\label{thm:mvcTerminate}
		Let $R$ be an Algorithm~\ref{alg:consensus}'s execution in which all correct nodes invoke $\mathsf{propose}()$ within $\bigO(1)$ CRWF/ACAF. BC-completion holds during $R$.
	\end{Theorem}	
	\renewcommand{\thmcnt}{\ref{thm:mvcTerminate}}
	\begin{theoremProof}
		We show: every correct node decides within $\bigO(1)$ CRWF/ACAF, \ie $\forall i \in \Correct: \done_i() \neq \bot$.
		
		\begin{Lemma}
			\label{thm:mainMVCTermB}
			Eventually, $\done_i()$ cannot return $\bot$ due to line~\ref{ln:notReady}.
		\end{Lemma}
		\renewcommand{\lemcnt}{\ref{thm:mainMVCTermB}}
		\begin{lemmaProof}
			Any correct node, $p_i$, asserts $\mathit{bcO}_i\neq \bot$, say, by invoking $\mathit{bcO}_i.\mathsf{propose}()$ (line~\ref{ln:bcOpropuse}).
			This is due to the assumption that all correct nodes invoke $\mathsf{propose}()$ within $\bigO(1)$ CRWF/ACAF, the definition of $\mathsf{propose}()$ (line~\ref{ln:mvcPropuseV}), VBB-completion, and the presence of at least $(n\mathit{-}t)$ correct nodes, which implies that $(\exists {S \subseteq\sP: n \mathit{-}t\leq |S|}): \forall {p_k \in S} : \mathsf{vbbDeliver}_i(k)\neq \bot$ holds within $\bigO(1)$ CRWF/ACAF, and the if-statement condition in line~\ref{ln:bcOpropuse} holds whenever $\mathit{bcO}_i= \bot$. 
			Eventually $\mathit{bcO}_i.\bcdone() \neq \bot$ (by the completion property of Binary consensus). 
			Thus, $\done_i()$ cannot return $\bot$ due to the if-statement in line~\ref{ln:notReady}.
		\end{lemmaProof}
		
		\smallskip
		
		If $\done_i()$ returns due to the if-statement in lines~\ref{ln:normalReturnAndConsis} to~\ref{ln:normalReturnAndConsisT}, $\done_i()\neq \bot$ is clear.
		Therefore, the rest of the proof focuses on showing that, within $\bigO(1)$ CRWF/ACAF, one of these three if-statement conditions must hold, and thus, the last return statement (of $\bot$ in line~\ref{ln:mvcDoneElad}) cannot occur, see Lemma~\ref{thm:mainMVCTerm}.
		
		\begin{Lemma}
			\label{thm:mainMVCTerm}
			Suppose lines~\ref{ln:normalReturnAndConsis} and~\ref{ln:normalReturnAndConsisT}'s conditions never hold \wrt any correct $p_i$.	
			Within $\bigO(1)$ CRWF/ACAF, line~\ref{ln:defultReturnIf}'s condition holds.
		\end{Lemma}
		\renewcommand{\lemcnt}{\ref{thm:mainMVCTerm}}
		\begin{lemmaProof}
			By VBB-completion, $\mathit{mcWait}_i()$ (line~\ref{ln:mcEchoExists}) must hold within $\bigO(1)$ CRWF/ACAF since there are $n-t$ correct and active nodes. 
			Thus, by the lemma assumption that the\reduce{ if-statement} condition in line~\ref{ln:normalReturnAndConsisT} never hold\reduce{ in $R$, we know that}, for any correct node $p_i$, $\True \in\mathit{bvO}_i.\binValues()$ holds within $\bigO(1)$ CRWF/ACAF, due to the properties of BV-broadcast (\SectionAbv~\ref{sec:sefBVbrodcast}).
			Thus, there is at least one correct node, $p_j$, for which $\sameValue_j()=\True$ when BV-broadcasting in line~\ref{ln:bcOpropuseTest}.
			By VBB-uniformity, the\reduce{ if-statement} condition in line~\ref{ln:defultReturnIf} must hold, within $\bigO(1)$ CRWF/ACAF, \wrt\reduce{ any correct node} $p_i$.
		\end{lemmaProof}
	\end{theoremProof}

	\Subsection{Closure of MVC}
	
	Theorem~\ref{thm:mvcClousre}'s proof shows that no consistency test causes false error indications.
	Theorem~\ref{thm:mvcClousre} counts communication rounds (without assuming fairness) using the CRWF notation (\SectionAbv~\ref{sec:CompletionVBB}).
	
	\begin{Theorem}[MVC closure]
		\label{thm:mvcClousre}
		Let $R$ be an Algorithm~\ref{alg:consensus}'s execution that starts in a post-recycling state and in which all correct nodes invoke $\mathsf{propose}()$ within $\bigO(1)$ CRWF.
		MVC requirements hold in $R$.
	\end{Theorem}
	
	\renewcommand{\thmcnt}{\ref{thm:mvcClousre}}
	\begin{theoremProof}
		BC-completion holds (Theorem~\ref{thm:mvcTerminate}). 
		\begin{Lemma}
			\label{thm:BCagreement}
			The BC-agreement property holds.
		\end{Lemma}	
		\renewcommand{\lemcnt}{\ref{thm:BCagreement}}
		\begin{lemmaProof}
			\emsA{We show} that no two correct nodes decide differently. 
			For every correct node, $p_i$, $\mathit{bcO}_i.\bcdone() \neq \bot$ holds \emsB{within $\bigO(1)$ CRWF} (Theorem~\ref{thm:mvcTerminate}). By the agreement and integrity properties of Binary consensus, $\mathit{bcO}_i.\bcdone() = \false$ implies BC-agreement (line~\ref{ln:normalReturnAndConsis}).  
			Suppose $\mathit{bcO}_i.\bcdone() = \true$. \emsA{The proof is implied since there is no correct node, $p_i$, and  (faulty or correct) node $p_k$} for which there is a value $w \notin \{\bot,\blitza,v\}$, such that $\mathsf{vbbDeliver}_i(k)=w$. This is due to $n \mathit{-} 2t \geq t + 1$ and $\sameValue()$'s second clause (line~\ref{ln:pbDef}), which requires $v$ to be unique.
		\end{lemmaProof}
		
		\begin{Lemma}
			\label{thm:MVC-validity}
			The BC-validity property holds.
		\end{Lemma}	
		\renewcommand{\lemcnt}{\ref{thm:MVC-validity}}
		\begin{lemmaProof}
			Suppose that all correct nodes propose the same value, $v$. The proof shows that $v$ is decided. Since all correct nodes propose $v$, we know that $v$ is validated (VBB-obligation). Also, all correct nodes VBB-deliver $v$ from at least $n \mathit{-} 2t$ different nodes (VBB-completion). Since $n \mathit{-} 2t > t$, value $v$ is unique.
			This is because no value $v'$ can be VBB-broadcast only by faulty nodes and still be validated (VBB-justification). Thus, the non-$\bot$ values that correct nodes can VBB-deliver are $v$ and $\blitza$. This means that $\forall i \in \Correct: \sameValue_i()=\true$, $\mathit{bcO}_i.\bcdone()=\true$ (Binary consensus validity), and all correct nodes decide $v$.
		\end{lemmaProof}
		
		\begin{Lemma}
			\label{thm:MVC-no-intrusion}
			The BC-no-intrusion property holds.
		\end{Lemma}	
		\renewcommand{\lemcnt}{\ref{thm:MVC-no-intrusion}}
		\begin{lemmaProof}
			Suppose $w\neq\blitza$ is proposed only by faulty nodes. The proof shows that no correct node decides $w$. By VBB-justification, no $p_i : i \in \Correct$ VBB-delivers $w$. 
			
			Suppose that $\mathit{bcO}_i.\bcdone() \neq \true$. Thus, $w$ is not decided due to\reduce{ the if-statement} line~\ref{ln:normalReturnAndConsis}. Suppose $\mathit{bcO}_i.\bcdone()=\true$. There must be a node $p_j$ for which $\sameValue_j()=\true$, 
			\ie $v$ is decided due to\reduce{ the if-statement in} line~\ref{ln:defultReturnIf} and since there are at least $n \mathit{-} 2t$ VBB-deliveries of $v$. 
			\ems{Note that the\reduce{ if-statement} condition in line~\ref{ln:normalReturnAndConsisT} cannot hold\reduce{ during $R$} since $R$ starts in a post-recycling system state as well as due to lines~\ref{ln:bcOpropuse} and~\ref{ln:bcOpropuseTest}, which use the same input value from $\sameValue()$.} 
			This implies that $w\neq v$ cannot be decided since $n \mathit{-} 2t > t$.
		\end{lemmaProof}
	\end{theoremProof}
	
} 
	
	\Section{Discussion}
	To the best of our knowledge, this paper presents the first \emsE{SSBFT MVC algorithm for} asynchronous message-passing systems. This solution is devised by layering SSBFT broadcast protocols. Our solution is based on a code transformation of existing (non-self-stabilizing) BFT algorithms into an SSBFT one. This transformation is achieved via careful analysis of the effect that arbitrary transient faults can have on the system's state as well as via rigorous proofs. We hope that the proposed solutions\reduce{ and studied techniques} can facilitate the design of new building blocks\reduce{, such as state-machine replication,} for the Cloud and distributed ledgers.
	         
	~~\\
	\textbf{Acknowledgments:} This work was partically supported by CyReV-2: Cyber Resilience for Vehicles - Cybersecurity for automotive systems in a changing environment (Vinnova 2019-03071).
	
	

\end{document}